\documentclass[twoside,12pt]{article}
\usepackage{epsfig}

\def\Journal#1#2#3#4{{#1} {#2} (#4) #3 }
\def\CMP{{\em Commun. Math. Phys.}}
\def\NPBPS{{\em Nucl. Phys.} B (Proc. Suppl.)}

\def\JHEP{{\em JHEP}}
\def\NCA{{\em Nuovo Cimento} A}
\def\PHYS{{\em Physica}}
\def\NPA{{\em Nucl. Phys.} A}

\def\EPJC{{\em Eur. Phys. J.} C}
\def\PRO{{\em Prog. Theor. Phys.}}
\def\NPB{{\em Nucl. Phys.} B}

\def\PLB{{\em Phys. Lett.} B}

\def\PRL{\em Phys. Rev. Lett.}
\def\PREV{\em Phys. Rev.}
\def\PREP{\em Phys. Rep.}

\def\PRD{{\em Phys. Rev.} D}
\def\PRC{{\em Phys. Rev.} C}

\def\ANNP{\em Ann. Phys. (N.Y.)}
\def\RMP{{\em Rev. Mod. Phys.}}

\def\IJMP{{\em Int. J. Mod. Phys.}}
\def\ARNPS{{\em Ann. Rev. Nucl.. Part. Sci.}}

\newcommand{\Z}{{\sf Z \!\!\! Z}}

\newcommand{\1}{{\sf 1 \!\! 1}}
\newcommand{\p}{\partial}
\newcommand{\Psibar}{\overline\Psi}
\newcommand{\psibar}{\overline\psi}
\newcommand{\chibar}{\overline\chi}

\begin{document}

\title{ \vspace{1cm} An Introduction to Chiral Symmetry on the Lattice}

\author{S.\ Chandrasekharan$^1$ and U.-J.\ Wiese$^2$ \\ \\
$^1$Department of Physics, Duke University, Durham, North Carolina, U.S.A.\\
$^2$Institute for Theoretical Physics, Bern University, Bern, Switzerland\\}

\maketitle

\begin{abstract}

The $SU(N_f)_L \otimes SU(N_f)_R$ chiral symmetry of QCD is of central 
importance for the nonperturbative low-energy dynamics of light quarks and 
gluons. Lattice field theory provides a theoretical framework in which these
dynamics can be studied from first principles. The implementation of chiral 
symmetry on the lattice is a nontrivial issue. In particular, local lattice 
fermion actions with the chiral symmetry of the continuum theory suffer from 
the fermion doubling problem. The Ginsparg-Wilson relation implies L\"uscher's
lattice variant of chiral symmetry which agrees with the usual one in the 
continuum limit. Local lattice fermion actions that obey the Ginsparg-Wilson 
relation have an exact chiral symmetry, the correct axial anomaly, they obey a 
lattice version of the Atiyah-Singer index theorem, and still they do not 
suffer from the notorious doubling problem. The Ginsparg-Wilson relation is 
satisfied exactly by Neuberger's overlap fermions which are a limit of Kaplan's
domain wall fermions, as well as by Hasenfratz and Niedermayer's classically 
perfect lattice fermion 
actions. When chiral symmetry is nonlinearly realized in effective field 
theories on the lattice, the doubling problem again does not arise. This review
provides an introduction to chiral symmetry on the lattice with an emphasis on 
the basic theoretical framework.

\end{abstract}
 
\maketitle
 
\newpage

\section{Introduction}

Physical phenomena arise over a vast range of energy scales. Attempts to 
unify gravity with the other fundamental forces suggest that the Planck scale
\begin{equation}
M_P = \frac{1}{\sqrt{G}} \approx 10^{19} \ \mbox{GeV},
\end{equation}
which is constructed from Newton's constant $G$ (and from $\hbar$ and $c$ which
we have put to 1) is the highest energy scale relevant to particle physics. On 
the other hand, ordinary matter receives almost all of its mass from protons 
and neutrons which have a mass $M \approx 1$ GeV. Can we understand why 
nucleons exist so far below the Planck scale? This is a typical hierarchy 
problem of which there are several in physics (also including the notorious 
cosmological constant problem). As Wilczek has pointed out, the large hierarchy
between the Planck scale $M_P$ and the nucleon mass $M$ is responsible for the 
feebleness of gravity \cite{Wil02}. To understand this, let us compare the 
strengths of the gravitational attraction and the electrostatic repulsion 
between two protons at some large distance $R$. The force of gravity is given 
by
\begin{equation}
F_g = G \frac{M^2}{R^2},
\end{equation}
while the electrostatic force is
\begin{equation}
F_e = \frac{e^2}{R^2},
\end{equation}
where $e$ is the proton's electric charge. The ratio of the two forces is thus
\begin{equation}
\frac{F_g}{F_e} = G \frac{M^2}{e^2} \approx 137 \frac{M^2}{M_P^2} \approx 
10^{-36}.
\end{equation}
Hence, if we can understand why $M \ll M_P$, we can understand why gravity is a
very weak force.

As Wilczek has explained in \cite{Wil02}, the nucleon mass $M$ is much smaller 
than the Planck scale $M_P$ partly due to the asymptotic freedom of QCD. At the
classical level, QCD with massless quarks has 
no dimensionful parameter at all. When the theory is quantized, a scale enters
through the mechanism of dimensional transmutation. A non-zero 
nucleon mass arises even in massless QCD due to the confinement of colored 
quarks and gluons inside color-neutral hadrons. Thus, the nucleon mass $M$ is a
nonperturbatively generated scale which cannot be understood using perturbation
theory. The continuum theory, i.e.\ dimensional regularization and 
renormalization applied to the QCD Lagrangian, is not even defined at a 
nonperturbative level. The only nonperturbative definition of QCD is provided 
by lattice field theory in which continuous space-time is replaced by a 
hypercubic lattice \cite{Wil74,Wil75}. In particular, it 
should be pointed out that lattice QCD is not an approximation to any 
pre-existing nonperturbatively well-defined theory in the continuum. Of course,
as in any other quantum field theory, one must ultimately remove the cut-off. 
On the lattice, the shortest physical distance is the lattice spacing $a$ which
defines an ultraviolet momentum cut-off $1/a$. Removing the cut-off thus means 
taking the continuum limit $a \rightarrow 0$. The masses of hadrons $M = 1/\xi$
are the inverse of a correlation length $\xi$. Taking the continuum limit means
that the physical mass $M$ must be much smaller than the cut-off, i.e.
\begin{equation}
M \ll \frac{1}{a} \ \Rightarrow \ \xi \gg a.
\end{equation}
Hence, in the continuum limit the physical correlation length $\xi$ goes to 
infinity in units of the lattice spacing. In the language of classical
statistical mechanics this corresponds to a second order phase transition.

Most of the time, lattice QCD is used as a very powerful tool for numerical
calculations of hadronic properties. However, the lattice can do more 
than that. To illustrate this, we will now use lattice QCD to explain why 
nucleons can exist naturally far below the Planck scale. Of course, it is 
well-known that QCD is not valid up to the Planck scale. In particular, it is 
embedded in the standard model which itself is an effective theory limited to 
energies below about 1 TeV. However, unlike the full standard model, thanks to 
asymptotic freedom QCD alone makes sense at arbitrarily high energy scales. 
Whatever replaces QCD and the standard model at ultra-short distances --- be it
string theory or some tiny wheels turning around at the Planck scale --- Nature
must have found a concrete way to regularize the QCD physics at ultra-short
distances. Due to renormalizability and universality, only the symmetries but 
not the details of this regularization should matter at low energies. For 
simplicity, we will use lattice QCD (and not, for example, string theory) as an
admittedly oversimplified model of Nature at ultra-short distances. In other
words, in this context we like to identify the lattice cut-off $1/a$ with the 
Planck scale $M_P$. 

Using lattice QCD, how can we then understand why the nucleon mass $M$ is far 
below $M_P = 1/a$? As Wilczek pointed out, one key ingredient is asymptotic 
freedom. Thanks to asymptotic freedom, without any fine-tuning of the bare 
gauge coupling a non-Abelian lattice Yang-Mills theory produces a correlation 
length $\xi$ that is larger than the lattice spacing $a$ by a factor 
exponentially large in the inverse coupling. In particular, choosing a bare 
coupling that is not unnaturally small, one can easily generate a hierarchy of 
scales like $M_P/M = \xi/a \approx 10^{19}$. Interestingly, the situation is 
not as simple when one proceeds from a pure gluon Yang-Mills theory to full 
lattice QCD including quarks. In particular, unlike continuum QCD, lattice QCD 
does not naturally have a chiral symmetry that can protect the quark masses 
from running up to the lattice ``Planck scale'' $1/a$. Indeed, for about two 
decades 
lattice field theorists have suffered from a hierarchy problem in the fermion 
sector. This problem first arose when Wilson removed the unwanted doubler 
fermions by breaking chiral symmetry explicitly \cite{Wil75}. Recovering chiral
symmetry in the continuum limit then requires a delicate fine-tuning of the 
bare fermion mass. In particular, if at ultra-short distances Nature would be a
lattice gauge theory with Wilson fermions, without unnatural fine-tuning
quarks would have masses at the Planck scale and the lightest particles would 
be glueballs. In that case it would be very puzzling why ordinary matter 
consists not just of gluons, but also of light quarks. If one works in 
continuum QCD one often takes chiral symmetry for granted, and one may view 
this hierarchy puzzle just as a problem of the lattice formulation. However,
one should not forget that continuum QCD is not even defined beyond 
perturbation theory. In addition, subtleties of the definition of $\gamma_5$ in
the framework of dimensional regularization affect even the continuum theory, 
and are just another aspect of the same deep problem of chiral symmetry that is
manifest on the lattice. Indeed, there is a severe hierarchy problem for 
nonperturbative fermion dynamics that Nature must have solved somehow because 
it presents us with nucleons that exist far below the Planck scale.

Remarkably, the long-standing hierarchy problem of lattice fermions --- and 
hence of the nonperturbative regularization of chiral symmetry --- has recently
been solved very elegantly. Using previous research of Callan and Harvey 
\cite{Cal85}, Kaplan \cite{Kap92} realized that massless four-dimensional
lattice fermions arise naturally, i.e.\ without fine-tuning, as zero-modes 
localized on a domain wall embedded in a five-dimensional space-time. In
particular, left- and right-handed fermions can be localized on a domain wall
and an anti-wall. When the wall and the anti-wall are separated by a 
sufficiently large distance, the left- and right-handed modes cannot mix, 
simply because they are spatially separated. As a result, a Dirac fermion 
arises which is protected from picking up a large mass and which is thus 
naturally light. Remarkably, in contrast to four dimensions, a Wilson term in a
five-dimensional lattice theory removes the doubler fermions without breaking 
the chiral symmetry of the light four-dimensional domain wall fermions.

When Kaplan proposed his idea of regulating chiral fermions using domain walls,
Narayanan and Neuberger were developing independently another approach to 
regulating chiral fermions using an infinite number of flavors 
\cite{Narayanan:wx}. Based on this approach they developed what
is now referred to as overlap lattice fermions \cite{Neu93,Narayanan:sk}. 
Since the flavor-space can be viewed as an extra dimension, the overlap 
approach is closely related to the domain wall approach. When one separates
the wall and the anti-wall by an infinite distance, domain wall fermions turn
into overlap fermions. Overlap fermions have the advantage that they have an 
exact chiral symmetry, while the chiral symmetry of domain wall fermions is
only approximate for a finite wall-anti-wall separation. Both overlap and
domain wall fermions yield naturally light quarks, and both are naturally 
related to the physics of a higher-dimensional space-time.

Hasenfratz and Niedermayer have investigated nonperturbative renormalization
group blocking transformations on the lattice \cite{Hasenfratz:1993sp}. The 
fixed points of such transformations correspond to lattice actions which 
are completely free of cut-off effects --- so-called perfect actions. 
Perfect actions for pure gauge theory, as well as for free Wilson and staggered
fermions were investigated in 
\cite{DeGrand:1995ji,Wie93,Bie96,Bietenholz:1996qc}. Classically perfect 
actions for full QCD have been constructed in 
\cite{Hasenfratz:1998jp,Hasenfratz:2000xz}. In the process of these 
investigations Hasenfratz (see \cite{Has98}) rediscovered an old paper by 
Ginsparg and Wilson \cite{Gin82}. He also realized that what is now called the 
Ginsparg-Wilson relation is the key to understanding chiral symmetry on the 
lattice. The Ginsparg-Wilson relation represents a general requirement on a 
lattice action which guarantees that it has good chiral properties. When 
Ginsparg and Wilson discovered this relation, it seemed impossible to 
explicitly construct lattice actions that obey it. By now it has been shown 
that classically perfect lattice actions can be approximated well enough, so 
that the Ginsparg-Wilson action is satisfied with high accuracy. From the point
of view of practical lattice QCD calculations this represents very important 
progress. However, the explicit construction of perfect actions is a delicate 
problem that can be considered a very elaborate form of fine-tuning. Hence, it 
seems unnatural that Nature has chosen anything like a perfect action to 
regularize the strong interactions at ultra-short distances. Unlike perfect 
fermions, overlap fermions can describe massless quarks in QCD without 
fine-tuning. By integrating out the extra dimension, Neuberger has constructed 
lattice Dirac operators for massless quarks analytically 
\cite{Neuberger:1997fp,Neu98a} and these Dirac operators do indeed 
satisfy the Ginsparg-Wilson relation exactly \cite{Neu98}. Remarkably, both
overlap as well as domain wall fermions, which naturally have a chiral symmetry
without fine-tuning, are related to the physics in a higher-dimensional 
space-time. Hence, the existence of light four-dimensional fermions may be a 
concrete hint to the physical reality of extra dimensions.

The full strength of the Ginsparg-Wilson relation was realized by L\"uscher
who discovered that it suggests a natural definition of lattice chiral symmetry
which reduces to the usual one in the continuum limit \cite{Lue98}. Based on
this insight, L\"uscher achieved a spectacular breakthrough: the 
nonperturbative construction of lattice chiral gauge theories \cite{Lue99}. 
Hence, not only QCD in which chiral symmetry is global, but also the standard
model with its local chiral symmetry now stands on a solid nonperturbative 
basis. Even continuum perturbation theory can benefit from these developments.
In particular, the ambiguities in the definition of $\gamma_5$  that arise in 
multi-loop calculations using dimensional regularization can be eliminated when
one uses the lattice regularization. Still, there is a very long way to go from
L\"uscher's theoretical construction to practical numerical calculations in 
chiral gauge theories like the standard model. 

The situation is a lot simpler, but still highly nontrivial, in applications of
Ginsparg-Wilson fermions to simulations of QCD. Compared to the standard Wilson
or staggered lattice fermions, which are already very difficult to treat fully 
dynamically, domain wall, overlap, or perfect fermions demand even much larger
computing power. Hence, at present they are often used in the quenched 
approximation in which the fermion determinant is ignored. If one does
not want to wait a long time for even bigger computers, it will require an
algorithmic breakthrough to bring the theoretical developments of lattice 
chiral symmetry to fruition in fully dynamical simulations of lattice QCD.
Promising steps in this direction are already being taken 
\cite{Hasenfratz:2002vv,Luscher:2003vf}.

If one imagines that Nature has used something like domain wall fermions to
regularize the strong interactions, it is natural that nucleons (and not just
glueballs) exist far below the Planck scale. However, it remains mysterious
where the quark masses themselves come from. In the standard model the quark 
masses arise from Yukawa couplings to the Higgs field, but the values of these
couplings are free parameters. Still, within the standard model the traditional
gauge hierarchy problem arises: why is the electroweak scale so small compared 
to the Planck scale? Chiral symmetry can protect fermion masses from running to
the ultimate high-energy cut-off, but it cannot protect scalars. A potential 
solution of the gauge hierarchy  problem is provided by supersymmetry.
Supersymmetry relates scalars to fermions and thus allows chiral symmetry to
indirectly generate naturally light scalars as well. At a nonperturbative
level, supersymmetry is as undefined as chiral symmetry was before the recent
developments on the lattice. In the worst case, supersymmetry may just be a
perturbative illusion which does not arise naturally at a nonperturbative 
level. Unfortunately, unlike for chiral symmetry, Nature has not yet provided 
us with experimental evidence for supersymmetry (except as an accidental
symmetry in heavy nuclei). Hence, one cannot be sure that it is indeed possible
to construct naturally light scalars at a nonperturbative level. Perhaps the
many beautiful results obtained within supersymmetric continuum theories should
make us optimistic that these theories actually exist at a rigorous level
beyond perturbation theory. Again, Kaplan and his collaborators have taken very
interesting steps towards constructing supersymmetric theories on the lattice
\cite{Kaplan:2003uh}. It remains to be seen if these developments will lead to 
a repetition of the Ginsparg-Wilson revolution of lattice chiral symmetry.

This review is an introduction to chiral symmetry on the lattice. We assume 
that the reader has a background in continuum field theory but not necessarily 
in lattice field theory. It should be noted that there are very good textbooks 
that cover the basics of lattice field theory much more completely than it 
can be done here \cite{Creutz,Muenster,Rothe:kp}. Furthermore, there are 
excellent reviews of the more advanced and specialized aspects of lattice 
field theory and chiral symmetry 
\cite{Gupta:1997nd,Luscher:1998pe,Creutz:2000bs,Neuberger:2001nb}. This 
review aims at bridging some of the gaps between the basic and the more 
advanced texts and introducing the reader to some of the latest developments 
in the field.
In section 2 we summarize symmetry properties of QCD in the continuum. We 
concentrate on chiral symmetry, but also discuss scale invariance, the axial
anomaly, and the Atiyah-Singer index theorem. Section 3 illustrates basic 
properties of lattice chiral symmetry using free fermions. In particular, we
discuss the fermion doubling problem, the Nielsen-Ninomiya theorem, and Wilson,
staggered, as well as perfect fermions. In section 4, gluons are added and 
full lattice QCD is considered. In particular, the Ginsparg-Wilson relation is
discussed. Section 5 concerns special features of Ginsparg-Wilson fermions, 
including the axial anomaly and the index theorem on the lattice, the 
Witten-Veneziano mass formula, the renormalization of operators, as well as
numerical simulations of Ginsparg-Wilson fermions. Section 6 discusses
effective theories for pions, nucleons, and constituent quarks, both in the
continuum and on the lattice. In particular, the nonlinear realization of 
chiral symmetry implies another way of solving the fermion doubling problem. 
Lattice simulations of constituent quarks may eventually shed some light on the
success of the nonrelativistic quark model. Finally, section 7 contains our
conclusions.

\section{Symmetries of the Strong Interactions}

In this section we review some aspects of symmetries in the continuum 
formulation of QCD with an emphasis on chiral symmetry.

\subsection{\it $SU(N_c)$ Yang-Mills Theory in the Continuum}

Let us consider an anti-Hermitean non-Abelian $SU(N_c)$ gauge field
\begin{equation}
A_\mu(x) = i g A^a_\mu(x) T^a,
\end{equation}
which (for $N_c = 3$) describes the gluons of QCD. Here $g$ is the gauge
coupling, $A^a_\mu(x)$ (with $a \in \{1,2,...,N_c^2 - 1\}$) is the real-valued 
non-Abelian vector potential at the Euclidean space-time point $x$, and the 
$T^a$ (which obey $\mbox{Tr}(T^a T^b) = \frac{1}{2} \delta_{ab}$) are the 
Hermitean generators of the $SU(N_c)$ algebra. The algebra-valued field 
strength takes the form
\begin{equation}
F_{\mu\nu}(x) = \p_\mu A_\nu(x) - \p_\nu A_\mu(x) + [A_\mu(x),A_\nu(x)],
\end{equation}
and the corresponding Euclidean Yang-Mills action is given by
\begin{equation}
S_{YM}[A] = - \int d^4x \ \frac{1}{2 g^2} \mbox{Tr}(F_{\mu\nu} F_{\mu\nu}).
\end{equation}
The action is invariant under group-valued gauge transformations 
$\Omega(x) \in SU(N_c)$,
\begin{equation}
A_\mu'(x) = \Omega(x) (A_\mu(x) + \p_\mu) \Omega(x)^\dagger,
\end{equation}
under which the field strength transforms as
\begin{equation}
F_{\mu\nu}(x) = \Omega(x) F_{\mu\nu}(x) \Omega(x)^\dagger.
\end{equation}
The quantum theory is defined by a functional integral over all gluon fields
\begin{equation}
Z = \int {\cal D}A \ \exp(- S_{YM}[A]),
\end{equation}
which is a formal expression before it is properly regularized. In perturbation
theory this is possible using standard dimensional regularization techniques. 
Through the regularization, a scale is introduced into the quantum 
theory which explicitly breaks the scale invariance of the classical Yang-Mills
theory. This anomaly in the scale invariance is responsible for the phenomenon 
of dimensional transmutation: in the quantum theory the dimensionless coupling 
constant $g$ of the classical theory is traded for a dimensionful scale. In the
modified minimal subtraction renormalization scheme this scale is 
$\Lambda_{\overline{MS}}$ which is defined in the framework of perturbation 
theory. We will soon define the theory beyond perturbation theory by 
regularizing it on a space-time lattice. In a nonperturbative context, a 
natural scale is the dynamically generated mass gap $M$ --- the energy of the 
lowest state above the vacuum. In a Yang-Mills theory this state is the 
lightest glueball. The $SU(N_c)$ Yang-Mills theory is a quantum theory without 
any free parameter. For example, the dimensionless ratio 
$M/\Lambda_{\overline{MS}}$ is a pure number predicted by the theory. The 
relation of $M$ or $\Lambda_{\overline{MS}}$ to units like GeV, on the other 
hand, is, of course, not predicted by the theory. Such man-made mass units are
related to the kilogram, defined by the arbitrary amount of platinum-iridium 
alloy deposited near Paris a long time ago.

Another quantity of physical interest is the topological charge
\begin{equation}
\label{topo}
Q[A] = - \frac{1}{32 \pi^2} \int d^4x \ \varepsilon_{\mu\nu\rho\sigma} 
\mbox{Tr}(F_{\mu\nu} F_{\rho\sigma}) \in \Pi_3[SU(N_c)] = \Z,
\end{equation}
which takes integer values in the third homotopy group of the gauge group. The
topological charge gives rise to an additional parameter, the vacuum angle
$\theta$, in the Yang-Mills functional integral
\begin{equation}
Z(\theta) = \int {\cal D}A \ \exp(- S_{YM}[A] + i \theta Q[A]).
\end{equation}
For $\theta \neq 0$ or $\pi$ the $\theta$-term explicitly breaks parity as well
as CP. The bound $|\theta| < 10^{-9}$ derived from the measurement of the 
electric dipole moment of the neutron suggests that $\theta = 0$ in Nature.  
This result is puzzling because in the Standard Model CP is already explicitly 
broken by the complex phase of the Cabbibo-Kobayashi-Maskawa matrix. The puzzle
to understand why $\theta = 0$ is known as the strong CP problem.

\subsection{\it QCD with $N_f$ Quark Flavors}

In the next step we add $N_f$ massless quarks to the pure gluon theory. Quarks
and anti-quarks are described by anti-commuting Dirac spinor fields $\psi(x)$
and $\psibar(x)$. In Euclidean space-time these two fields represent 
independent Grassmann degrees of freedom. Under a non-Abelian gauge 
transformation the quark and anti-quark fields transform in the fundamental
representations $\{N_c\}$ and $\{\overline{N_c}\}$, respectively, i.e.
\begin{equation}
\psi(x)' = \Omega(x) \psi(x), \ \psibar(x)' = \psibar(x) \Omega(x)^\dagger.
\end{equation}
The fermionic part of the Euclidean action of massless QCD takes the form
\begin{equation}
S_F[\psibar,\psi,A] = \int d^4x \ \psibar \gamma_\mu (\p_\mu + A_\mu) \psi,
\end{equation}
which is gauge invariant by construction. The Euclidean Dirac matrices are 
Hermitean and obey the anti-commutation relations
\begin{equation}
\{\gamma_\mu,\gamma_\nu\} = 2 \delta_{\mu\nu}, \
\{\gamma_\mu,\gamma_5\} = 0, \
\gamma_5 = \gamma_1 \gamma_2 \gamma_3 \gamma_4.
\end{equation}
We now decompose the quark fields into left- and right-handed components
\begin{eqnarray}
&&\psi_L(x) = P_L \psi(x), \ \psi_R(x) = P_R \psi(x), \ 
\psi(x) = \psi_L(x) + \psi_R(x), \nonumber \\
&&\psibar_L(x) = \psibar(x) P_R, \ \psibar_R(x) = \psibar(x) P_L, \
\psibar(x) = \psibar_L(x) + \psibar_R(x).
\end{eqnarray}
The chiral projectors are given by
\begin{equation}
\label{projectors}
P_R = \frac{1 + \gamma_5}{2}, \ P_L = \frac{1 - \gamma_5}{2}.
\end{equation}
Inserting the decomposed spinors into the fermionic part of the action one 
obtains
\begin{equation}
S_F[\psibar,\psi,A] = \int d^4x \ 
\left[\psibar_L \gamma_\mu (\p_\mu + A_\mu) \psi_L +
\psibar_R \gamma_\mu (\p_\mu + A_\mu) \psi_R \right],
\end{equation}
i.e.\ the action decouples into two contributions from left- and right-handed 
quarks. 

As a result, the action of massless QCD is invariant against $U(N_f)_L \otimes
U(N_f)_R$ chiral transformations
\begin{eqnarray}
&&\psi'_L(x) = L \ \psi_L(x), \ \psibar'(x) = \psibar_L(x) L^+, \ L \in
U(N_f)_L, \nonumber \\
&&\psi'_R(x) = R \ \psi_R(x), \ \psibar'(x) = \psibar_R(x) R^+, \ R \in 
U(N_f)_R.
\end{eqnarray}
Due to an anomaly in the axial $U(1)_A$ symmetry, the symmetry of the quantum 
theory is reduced to $SU(N_f)_L \otimes SU(N_f)_R \otimes U(1)_B$ where the 
$U(1)_B = U(1)_{L=R}$ symmetry represents baryon number conservation.

Chiral symmetry is only approximate in Nature, because the quark mass terms
couple left- and right-handed fermions. The mass terms in the QCD action take 
the form
\begin{equation}
S_M[\psibar,\psi] = \int d^4x \ 
\left[\psibar_R {\cal M} \psi_L + \psibar_L {\cal M}^\dagger \psi_R \right],
\end{equation}
which is again gauge invariant but no longer chirally invariant. The quark 
mass matrix takes the form
\begin{equation}
{\cal M} = \mbox{diag}(m_u,m_d,m_s,...,m_{N_f}).
\end{equation}
If all quark masses are equal, i.e.\ if ${\cal M} = m \1$, the mass term is 
invariant only against simultaneous transformations $L = R$. Hence, chiral 
symmetry is then explicitly broken down to 
\begin{equation}
SU(N_f)_{L=R} \otimes U(1)_{L=R} = SU(N_f)_F \otimes U(1)_B,
\end{equation}
which corresponds to the flavor and baryon number symmetry. In Nature the
quark masses are all different, and the symmetry is, in fact, explicitly 
broken down to
\begin{equation}
\prod_{f=1}^{N_f} U(1)_f = U(1)_u \otimes U(1)_d \otimes U(1)_s \otimes ...
\otimes U(1)_{N_f}.
\end{equation}
The physical up and down quark masses are a lot smaller than 
$\Lambda_{\overline{MS}}$ while the strange quark mass is of the order of 
$\Lambda_{\overline{MS}}$. Consequently, $SU(2)_L \otimes SU(2)_R$ is a very
good approximate global symmetry, while $SU(3)_L \otimes SU(3)_R$ is broken
more strongly. It should be noted that the actual values of the quark 
masses are reasonably well-known from comparison with experiment, but are at 
present not at all understood theoretically. In particular, we don't know why 
there are three light quark flavors. Before one understands the relevant 
physics beyond the Standard Model, the origin of the chiral symmetry of QCD 
remains mysterious and the symmetry itself seems accidental.

The total action of QCD is simply given by
\begin{equation}
S_{QCD}[\psibar,\psi,A] = 
S_{YM}[A] + S_F[\psibar,\psi,A] + S_M[\psibar,\psi],
\end{equation}
and the corresponding QCD functional integral is
\begin{equation}
Z = \int {\cal D}\psibar {\cal D}\psi {\cal D}A \ 
\exp(- S_{QCD}[\psibar,\psi,A]).
\end{equation}
Again, this is a formal mathematical expression before it is properly 
regularized. In the continuum this can be done only perturbatively. We will 
soon discuss the lattice regularization which defines QCD beyond perturbation
theory.

\subsection{\it The Axial Anomaly and the Atiyah-Singer Index Theorem}

The $U(1)_A$ symmetry of the classical action of massless QCD is explicitly
broken by quantum effects. As a consequence of this anomaly the flavor-singlet
axial current
\begin{equation}
j_\mu^5(x) = \psibar(x) \gamma_\mu \gamma_5 \psi(x),
\end{equation}
which is conserved at the classical level, has a non-zero divergence
\begin{equation}
\p_\mu j_\mu^5(x) = - \frac{N_f}{32 \pi^2} \varepsilon_{\mu\nu\rho\sigma} 
\mbox{Tr}[F_{\mu\nu}(x) F_{\rho\sigma}(x)],
\end{equation}
due to instantons (and other topological charge carriers) in the quantum 
theory. In particular, the variation of the axial charge 
$Q^5(t) = \int d^3x \ j_0(\vec x,t)$ is given by
\begin{equation}
Q^5(t = \infty) - Q^5(t = - \infty) = N_f \ Q[A],
\end{equation}
where $Q[A]$ is the topological charge of eq.(\ref{topo}). Only in the 
$N_c \rightarrow \infty$ limit the anomaly vanishes and the chiral symmetry of 
massless QCD is enhanced to the full $U(N_f)_L \otimes U(N_f)_R$ group.

The axial anomaly is deeply connected with the Atiyah-Singer index theorem, 
which relates the zero-modes of the massless ($N_f$-flavor) Dirac operator 
$D[A] = \gamma_\mu (\p_\mu + A_\mu)$ to the topological charge $Q[A]$. 
The eigenvalues of the Dirac operator are purely imaginary and come in complex 
conjugate pairs. Only the zero eigenvalues are not paired. Since the Dirac
operator anti-commutes with $\gamma_5$, the eigenvectors of the zero-modes
(which obey $D[A] \psi = 0$) have a definite handedness, i.e.\ 
$\gamma_5 \psi = \pm \psi$. The index theorem states that 
\begin{equation}
\mbox{index}(D[A]) = n_- - n_+ = N_f \ Q[A],
\end{equation}
i.e.\ the index of the operator $D[A]$, which is defined as the difference 
between the number of left- and right-handed zero-modes, is given by the 
topological charge. 

As a consequence of the index theorem, topologically nontrivial gluon field 
configurations (with $Q[A] \neq 0$) necessarily induce zero-modes in the 
Dirac operator and thus lead to a vanishing fermion determinant 
$\mbox{det} D[A] = 0$. As a function of the vacuum angle $\theta$, the 
functional integral of massless QCD takes the form
\begin{eqnarray}
Z(\theta)&=&\int {\cal D}\psibar {\cal D}\psi {\cal D}A \ 
\exp(- S_{QCD}[\psibar,\psi,A] + i \theta Q[A]) \nonumber \\
&=&\int {\cal D}A \ \mbox{det} D[A] \exp(- S_{YM}[A] + 
i \theta Q[A]) \nonumber \\
&=&\int {\cal D}A \ \mbox{det} D[A] \exp(- S_{YM}[A]) = Z(0). 
\end{eqnarray}
Since $\mbox{det} D[A] = 0$ when $Q[A] \neq 0$, there are no 
$\theta$-vacuum effects in massless QCD. This would ``solve'' the strong CP 
problem (why is $\theta = 0$ ?) if, for example, the up quark would be 
massless. Of course, this would leave us with the ``up quark problem'': why 
should $m_u = 0$ ? In any case, $m_u = 0$ seems not to be realized in Nature 
and the strong CP problem remains puzzling.

\subsection{\it Spontaneous Chiral Symmetry Breaking}

Due to the approximate chiral symmetry of QCD one would expect corresponding 
near degeneracies in the spectrum of strongly interacting particles. Indeed, 
hadrons can be classified as isospin multiplets. The isospin transformations 
act on left- and right-handed fermions simultaneously, i.e.\ 
$SU(2)_I = SU(2)_{L=R}$. The $SU(3)_F = SU(3)_{L=R}$ flavor symmetry is more 
approximate but is still clearly visible in the spectrum. The full 
$SU(N_f)_L \otimes SU(N_f)_R \otimes U(1)_B$ chiral symmetry, on the other 
hand, is not manifest in the spectrum at all. In particular, one does not 
observe mass-degenerate parity doublets of hadrons, as one should if chiral 
symmetry was manifest in the spectrum. Furthermore, one observes very light 
pseudo-scalar particles --- the pions $\pi^+$, $\pi^0$, and $\pi^-$ --- as well
as somewhat heavier pseudo-scalars --- the four kaons $K^+$, $K^0$, 
$\overline{K^0}$, $K^-$ and the $\eta$-meson.

>From the experimental evidence one concludes that chiral symmetry must be
spontaneously broken. Indeed, when a continuous global symmetry breaks 
spontaneously, massless Goldstone bosons appear in the spectrum. According to 
Goldstone's theorem, the number of massless bosons is given by the difference 
of the number of generators of the full symmetry group $G$ and the subgroup $H$
that remains unbroken. In massless QCD the full chiral symmetry group is
\begin{equation}
G = SU(N_f)_L \otimes SU(N_f)_R \otimes U(1)_B,
\end {equation}
while the unbroken subgroup is the flavor symmetry
\begin{equation}
H = SU(N_f)_{L=R} \otimes U(1)_B.
\end{equation}
Hence, in this case one expects $N_f^2 - 1$ massless Goldstone bosons. For 
$N_f = 2$ these are the three pions, while for $N_f = 3$ there are eight 
Goldstone bosons --- the pions, the kaons, and the $\eta$-meson. In Nature 
these particles are not exactly massless, because chiral symmetry is explicitly
broken by the quark masses. The masses of the up and down quarks are much 
smaller than the QCD scale $\Lambda_{\overline{MS}}$ which leads to the very 
small pion mass. The mass of the strange quark, on the other hand, is of the 
order of $\Lambda_{\overline{MS}}$, thus leading to larger masses of the kaons 
and the $\eta$-meson. Still, their masses are small enough to identify these 
particles as pseudo-Goldstone bosons.

Chiral symmetry breaking has not yet been derived analytically from the QCD 
Lagrangian. In particular, spontaneous chiral symmetry breaking is a 
nonperturbative phenomenon whose understanding requires a formulation of QCD 
beyond perturbation theory. Such a formulation is provided by lattice field 
theory which will be discussed below. In the lattice formulation, in the
strong coupling limit it is possible to show rigorously that chiral symmetry 
is indeed spontaneously broken \cite{Salm91,Salm91a}. Further, numerical 
simulations in lattice QCD confirm that chiral symmetry is spontaneously 
broken even at weaker couplings. For example, one 
detects spontaneous chiral symmetry breaking by investigating the chiral order 
parameter
\begin{equation}
\langle \psibar \psi \rangle = \langle 0|\psibar(x)\psi(x)|0 \rangle =
\langle 0|\psibar_R(x) \psi_L(x) + \psibar_L(x) \psi_R(x)|0 \rangle.
\end{equation}
The order parameter is invariant against simultaneous transformations $R = L$,
but not against general chiral rotations. If chiral symmetry would be intact
the chiral condensate would vanish. When the symmetry is spontaneously broken,
on the other hand, $\langle \psibar \psi \rangle$ is non-zero.

\section{Free Lattice Fermions}

In this section we begin to formulate QCD on a space-time lattice which serves 
as an ultraviolet regulator. We replace Euclidean space-time by a hypercubic 
lattice of points $x$ with lattice spacing $a$. The lattice provides an 
ultraviolet momentum cut-off $1/a$. The continuum limit is reached when 
$a \rightarrow 0$. One has a lot of freedom in writing down a lattice 
regularized theory. In order to ensure that one reaches the desired theory in 
the continuum limit one must pay attention to the relevant symmetries. The 
most important symmetry of QCD is the $SU(N_c)$ gauge invariance. It is an
important strength of the lattice regularization that it manifestly respects 
gauge invariance. The fact that space-time symmetries are explicitly broken 
down to discrete translations and the hypercubic rotation group of the lattice 
is not a severe problem. In particular, in QCD the hypercubic symmetry is 
powerful enough to ensure that the full Poincar\'e symmetry of the continuum is
automatically recovered as $a \rightarrow 0$. Discrete symmetries like parity 
and charge conjugation are also easy to maintain on the lattice. This review 
concentrates on the question of how to realize chiral symmetry on the lattice. 
In this section we consider lattice theories of free quarks only. Gluon fields 
will be added in the next section.

\subsection{\it The Naive Lattice Fermion Action and the Doubling Problem}

In the continuum the Euclidean action of a free Dirac fermion in $d$ space-time
dimensions is given by
\begin{equation}
S[\psibar,\psi] = \int d^dx \ \psibar (\gamma_\mu \partial_\mu + m) \psi,
\end{equation}
and the functional integral takes the form
\begin{equation}
Z = \int {\cal D}\psibar {\cal D}\psi \ \exp(- S[\psibar,\psi]).
\end{equation}
On the lattice the continuum fermion field $\psibar(x), \psi(x)$ is replaced by
Grassmann variables $\Psibar_x, \Psi_x$ which live on the lattice points $x$. 
The continuum derivative can be discretized by a finite difference, such that
\begin{equation}
S[\Psibar,\Psi] = a^d \ \Psibar D \Psi = a^d \sum_{x,\mu} \frac{1}{2a}
(\Psibar_x \gamma_\mu \Psi_{x+\hat\mu} - \Psibar_{x+\hat\mu} \gamma_\mu
\Psi_x) + a^d \sum_x m \Psibar_x \Psi_x.
\end{equation}
Here $\hat\mu$ is a vector of length $a$ in the $\mu$-direction. In the 
continuum limit $a \rightarrow 0$ the lattice sum $a^d \sum_x$ becomes the 
continuum integral $\int d^dx$ over space-time. The corresponding lattice Dirac
operator which is a matrix in the Dirac- and space-time indices takes the form
\begin{equation}
D_{x,y} = \sum_\mu \frac{1}{2a} (\gamma_\mu \delta_{x+\hat\mu,y} -
\gamma_\mu \delta_{x-\hat\mu,y}) + m \delta_{x,y}.
\end{equation}
The lattice functional integral can be written as
\begin{equation}
Z = \int {\cal D}\Psibar {\cal D}\Psi \ \exp( -a^d \ \Psibar D \Psi) =
\prod_x \int d\Psibar_x d\Psi_x \ \exp(- S[\Psibar,\Psi]).
\end{equation}
In particular, the fermionic Grassmann integration measure is completely 
regularized explicitly.

The momentum space of the lattice theory is a $d$-dimensional Brillouin zone 
$B = [- \pi/a,\pi/a]^d$ with periodic boundary conditions. Going to momentum 
space, the naive fermion action from above gives rise to the lattice fermion 
propagator
\begin{equation}
\label{prop}
\langle \Psibar(-p) \Psi(p) \rangle = 
[i \sum_\mu \gamma_\mu \frac{1}{a} \sin(p_\mu a) + m]^{-1}.
\end{equation}
By performing a Fourier transform in the Euclidean energy $p_d$ one obtains the
fermion 2-point function
\begin{equation}
\langle \Psibar(- \vec p,0) \Psi(\vec p,x_d) \rangle = \frac{1}{2 \pi}
\int_{- \pi/a}^{\pi/a} dp_d \ \langle \Psibar(-p) \Psi(p) \rangle 
\exp(i p_d x_d) \sim \exp(- E(\vec p) x_d).
\end{equation}
At large Euclidean time separation $x_d$ the 2-point function decays 
exponentially with the energy $E(\vec p)$ of a fermion with spatial momentum 
$\vec p$. For the naive fermion action the lattice dispersion relation takes 
the form
\begin{equation}
\sinh^2(E(\vec p) a) = \sum_i \sin^2(p_i a) + (m a)^2.
\end{equation}
The continuum dispersion relation $E(\vec p)^2 = \vec p \ ^2 + m^2$ is indeed
recovered in the continuum limit $a \rightarrow 0$. However, besides 
$\vec p = 0$ there are other momenta $\vec p$ for which $E(\vec p)$ becomes 
small. These are located at the corners of the Brillouin zone where the
components of the momentum vector take the values $p_i = 0$ or $\pi/a$, such 
that $\sin(p_i a) = 0$. As a consequence, the lattice dispersion relation leads
to additional states in the spectrum which are absent in the continuum theory 
and which do not disappear in the continuum limit. Hence, the naive lattice 
fermion action does not lead to the correct continuum theory. The extra states 
appearing in the lattice dispersion relation show up as additional physical 
particles --- the so-called doubler fermions. Fermion doubling is a 
manifestation of a deep fundamental problem of lattice regularized fermionic 
theories with a chiral symmetry. The fermion doubling problem leads to a 
multiplication of fermion species. The lattice fermion propagator of 
eq.(\ref{prop}) has $2^d$ poles instead of just one as in the continuum. The 
origin of the doubling problem is deeply connected with chiral symmetry and can
be traced back to the axial anomaly. The doubler fermions pose a severe problem
in lattice field theory. Without removing them we cannot describe Nature's QCD 
(which has 3 and not $2^d = 2^4 = 16$ light quark flavors).

\subsection{\it The Nielsen-Ninomiya Theorem}

Before we try to eliminate the doubler fermions let us prove a general theorem
due to Nielsen and Ninomiya \cite{Nie81}: a chirally invariant free fermion 
lattice action, which is local, translation invariant, and real necessarily 
has fermion doubling. The theorem is based on topology. It holds because the 
lattice momentum space (the Brillouin zone $B$) is a torus. A general chirally
symmetric and translationally invariant lattice action for free fermions takes 
the form
\begin{equation}
S[\Psibar,\Psi] = a^d \sum_{x,y} \Psibar_x \gamma_\mu \rho_\mu(x-y) \Psi_y.
\end{equation}
The function $\rho_\mu(x-y)$ determines the strength of the coupling between
the fermion field values $\Psibar_x$ and $\Psi_y$ at two points $x$ and $y$ 
which may be 
separated by an arbitrarily large distance. Locality of the lattice action does
not mean that the points $x$ and $y$ must be nearest neighbors. It only means 
that $\rho_\mu(x-y)$ decays exponentially at large separations $x-y$. Going to 
momentum space, locality implies that in Fourier space $\rho_\mu(p)$ is a
regular function (without poles) over the Brillouin zone. The corresponding 
lattice fermion propagator takes the form
\begin{equation}
\langle \Psibar(-p) \Psi(p) \rangle =
[i \sum_\mu \gamma_\mu \rho_\mu(p)]^{-1}.
\end{equation}
Reality and translation invariance of the lattice action imply that 
$\rho_\mu(p)$ is a real-valued periodic function over the Brillouin zone.

Poles of the propagator --- and hence physical or doubler fermions ---
correspond to zeros of $\rho_\mu(p)$, i.e.\ to points $p$ with 
$\rho_\mu(p) = 0$ for all $\mu$. The Nielsen-Ninomiya theorem states that a
regular, real-valued, and periodic function $\rho_\mu(p)$ necessarily vanishes 
at more than just one point. It is trivial to prove this for $d = 1$. In that
case, there is a single regular periodic function $\rho_1(p)$ which should at 
least have one zero in order to describe the physical fermion pole. The 
function is positive on one side of the zero and negative on the other side. 
Hence, it must go through zero again in order to satisfy periodicity, thus 
leading to a doubler fermion pole in the lattice propagator. In higher 
dimensions the proof is analogous. For example, for $d = 2$ there are two 
functions $\rho_1(p)$ and $\rho_2(p)$. The zeros of $\rho_1(p)$ lie on a closed
curve in the two-dimensional Brillouin zone. This curve may be closed via the 
periodic boundary conditions. The zeros of $\rho_2(p)$ lie on another closed 
curve that intersects the first one in the pole position of the physical 
fermion. Due to the periodic boundary conditions of the Brillouin zone, the two
curves must necessarily also intersect somewhere else. The curves cannot just 
touch each other because this would lead to an incorrect dispersion relation 
for the physical fermion. In $d$ dimensions the zeros of $\rho_\mu(p)$ (with 
$\mu = 1,2,...,d$) lie on $d$ closed $(d-1)$-dimensional surfaces. Again, those
cannot intersect in just one point. If they intersect once they necessarily 
intersect also somewhere else. This proves lattice fermion doubling for a 
chirally symmetric, translation invariant, real-valued lattice action. It 
should be noted that the theorem does not specify the number of doubler 
fermions. It is indeed possible to reduce the number of doublers from $2^d-1$ 
to 1, but it is impossible to eliminate the doubler fermions completely. 

One may try to evade the theorem by violating one of its basic assumptions.
Giving up translation invariance or the reality of the action has not led to 
acceptable solutions of the fermion doubling problem. Giving up locality is 
probably the last thing one should do in field theory. For example, the early 
idea of SLAC fermions \cite{SLAC} turned out to be unacceptable for this 
reason. 

\subsection{\it Wilson Fermions}

In his work on lattice gauge theory Wilson removed the doubler fermions in a
direct and radical way by breaking chiral symmetry explicitly \cite{Wil75}. 
Then the Nielsen-Ninomiya theorem is evaded because the propagator contains 
additional 
terms without $\gamma_\mu$. The so-called Wilson term gives the fermion 
doublers a mass of the order of the cut-off while the physical fermion remains 
massless. Hence, in the continuum limit chiral symmetry is recovered in the 
physical sector. Wilson's modification of the naive fermion action takes the 
form of a discretized second derivative
\begin{eqnarray}
\label{freewilson}
S[\Psibar,\Psi]&=&a^d \sum_{x,\mu} \frac{1}{2 a}(\Psibar_x \gamma_\mu 
\Psi_{x+\hat\mu} - \Psibar_{x+\hat\mu} \gamma_\mu \Psi_x) + 
a^d \sum_x m \Psibar_x \Psi_x \nonumber \\
&+& a^d \sum_{x,\mu} \frac{1}{2a} (2 \Psibar_x \Psi_x 
- \Psibar_x \Psi_{x+\hat\mu} - \Psibar_{x+\hat\mu} \Psi_x).
\end{eqnarray}
Then the lattice propagator takes the form
\begin{equation}
\langle \Psibar(-p) \Psi(p)\rangle = 
[i \sum_\mu \gamma_\mu \frac{1}{a} \sin(p_\mu a) + m + 
\sum_\mu \frac{2}{a} \sin^2(\frac{p_\mu a}{2})]^{-1}.
\end{equation}
The Wilson term acts as a momentum-dependent mass term. For small momenta it 
vanishes quadratically, and hence it does not affect the dispersion of the
physical fermion, at least in the continuum limit. For the doubler fermions,
on the other hand, the Wilson term is non-zero, and gives them a mass of the 
order of the cut-off $1/a$. In the continuum limit the doubler fermions are 
hence eliminated from the spectrum of the theory. Unfortunately, in lattice QCD
this leads to a variety of complications. In particular, recovering chiral 
symmetry in the continuum limit requires unnatural fine-tuning of the bare 
fermion mass. For more details we refer the reader to \cite{Muenster,Rothe:kp}.

\subsection{\it Staggered Fermions}

Staggered fermions result from naive doubled lattice fermions by so-called spin
diagonalization 
\cite{Susskind:1976jm,Sharatchandra:1981si,Kluberg-Stern:1983dg}. 
In four space-time dimensions the size of the Dirac matrices 
is $4 \times 4$. By spin diagonalization one can reduce the fermion 
multiplication factor $2^d = 16$ to $16/4 = 4$. Hence, for $d = 4$, staggered 
fermions represent 4 flavors of mass-degenerate fermions. Staggered fermions 
have a single 1-component pair of Grassmann variables $\chibar_x$ and $\chi_x$ 
per lattice point $x$. The corresponding lattice action for free 
staggered fermions takes the form
\begin{equation}
S[\chibar,\chi] = 
a^4 \sum_{x,\mu} \frac{1}{2a} 
(\chibar_x \eta_{x,\mu} \chi_{x+\hat\mu} - 
\chibar_{x+\hat\mu} \eta_{x,\mu} \chi_x) + a^4 \sum_x m \chibar_x \chi_x ,
\end{equation}
where
\begin{equation}
\eta_{x,1} = 1, \ \eta_{x,2} = (-1)^{x_1/a}, \ 
\eta_{x,3} = (-1)^{(x_1+x_2)/a}, \ \eta_{x,4} = (-1)^{(x_1+x_2+x_3)/a}.
\end{equation}
For $m = 0$ the staggered fermion action has an exact $U(1)_e \otimes U(1)_o$ 
symmetry
\begin{eqnarray}
\chi'_x = \exp(i \varphi_e) \chi_x, \ 
\chibar'_x = \chibar_x \exp(- i \varphi_o), \ 
\mbox{for} \ (x_1 + x_2 + x_3 + x_4)/a \ \mbox{even}, \nonumber \\
\chi'_x = \exp(i \varphi_o) \chi_x, \ 
\chibar'_x = \chibar \exp(- i \varphi_e), \ 
\mbox{for} \ (x_1 + x_2 + x_3 + x_4)/a \ \mbox{odd},
\end{eqnarray}
which is a subgroup of the $SU(4)_L \otimes SU(4)_R \otimes U(1)_B$ chiral 
symmetry of the corresponding continuum theory. In the interacting theory the 
chiral and flavor symmetries besides $U(1)_e \otimes U(1)_o$ are explicitly 
broken by the staggered fermion action. These symmetries are recovered only in 
the continuum limit. A detailed introduction to staggered fermions and its
properties can be found in textbooks like \cite{Muenster,Rothe:kp} and 
review articles like \cite{Kilcup:1986dg} which cover more advanced topics.

Unlike Nature's QCD, staggered fermions have 4 flavors of the same mass. Since 
they have a remnant of chiral symmetry and are relatively easy to simulate 
numerically, staggered fermions provide a convenient framework for studies of 
chiral symmetry breaking at $N_f = 4$. Some numerical studies of staggered 
fermions try to mimic Nature's QCD by taking roots of the fermion determinant 
in order to relate 4 mass-degenerate staggered fermions to physical up, down, 
and strange quarks. These calculations are not on completely solid grounds 
before one can show that locality (which may be violated at finite lattice 
spacing) is recovered in the continuum limit.

\subsection{\it Perfect Lattice Fermions}

In this subsection we relate the continuum theory of free fermions to a 
corresponding lattice theory by an exact renormalization group transformation.
This is achieved by defining lattice fermion fields as block averages of
continuum fields integrated over hypercubes 
\cite{Wie93,Bie96,Bietenholz:1996qc}. The resulting lattice theory is in
all respects equivalent to the underlying continuum theory, i.e.\ it is 
completely free of lattice artifacts. For example, it has the same
energy-momentum dispersion relation as the continuum theory. Even more 
important, it has an exact chiral symmetry (which may, however, be hidden).
Lattice actions with these properties are known as perfect actions.

Let us derive a perfect fermion action by blocking from the continuum
\cite{Bie96}. For this purpose we average the continuum fermion field $\psi(y)$
over hypercubes $c_x$ of size $a^d$ centered at the points $x$ of a 
$d$-dimensional Euclidean lattice
\begin{equation}
\Psi_x = \frac{1}{a^d} \int_{c_x} d^dy \ \psi(y), \
\Psibar_x = \frac{1}{a^d} \int_{c_x} d^dy \ \psibar(y),
\end{equation}
which in momentum space corresponds to
\begin{equation}
\Psi(p) = \sum_{l \in \Z^d} \psi(p + 2 \pi l/a) \Pi(p + 2 \pi l/a), \ 
\Psibar(-p) = \sum_{n \in \Z^d} \psibar(- p - 2 \pi n/a) \Pi(p + 2 \pi n/a), \ 
\end{equation}
Note that the lattice fermion field is periodic over the Brillouin zone. The 
Fourier transform of the blocking kernel is given by 
\begin{equation}
\Pi(p) = \prod_{\mu = 1}^d \frac{2 \sin(p_\mu a/2)}{p_\mu a}.
\end{equation}
The lattice fermion propagator is related to the continuum propagator by 
\begin{eqnarray}
\label{perfprop}
\langle \Psibar(-p) \Psi(p) \rangle&=& 
\sum_{l \in \Z^d} \langle \psibar(- p - 2 \pi l/a) 
\psi(p + 2 \pi l/a) \rangle \Pi(p + 2 \pi l/a)^2 \nonumber \\
&=&\sum_{l \in \Z^d} [i \gamma_\mu (p_\mu + 2 \pi l_\mu/a) + m]^{-1} 
\Pi(p + 2 \pi l/a)^2.
\end{eqnarray}
For $m = 0$ the lattice propagator corresponds to a lattice action
\begin{equation}
S[\Psibar,\Psi] = a^d \sum_{x,y} \Psibar_x \gamma_\mu \rho_\mu(x-y) \Psi_y,
\end{equation}
with couplings $\rho_\mu(x-y)$ calculable by a Fourier transformation. This
lattice action is perfect by construction, i.e.\ its spectrum is identical with
the one of the continuum theory. Hence, there should be no fermion doubling.
On the other hand, the action is manifestly chirally invariant. This seems to
contradict the Nielsen-Ninomiya theorem. However, the theorem is evaded because
the action turns out to be nonlocal. Its couplings $\rho_\mu(x-y)$ do not decay
exponentially at large distances. Instead for $d \geq 2$ they decay only 
power-like. As a consequence, in momentum space $\rho_\mu(p)$ is not regular 
(it actually has poles) and therefore the topological arguments behind the 
Nielsen-Ninomiya theorem do not apply. The nonlocality can be seen easily for
$d=1$. Then $\gamma_1 = 1$ and the sum in eq.(\ref{perfprop}) can be performed 
analytically, resulting in a massless propagator that takes the form 
\begin{equation}
\langle \Psibar(-p) \Psi(p) \rangle = 
\sum_{l \in \Z} [i (p + 2 \pi l/a)]^{-1} 
\Pi(p + 2 \pi l/a)^2 = \frac{a}{2i} \cot(\frac{p a}{2}).
\end{equation}
This implies
\begin{equation}
\rho_1(p) = \frac{2}{a} \tan(\frac{p a}{2}),
\end{equation}
which is singular at the edge of the Brillouin zone ($p = \pm \pi/a$). The
corresponding coupling in coordinate space,
\begin{equation}
\rho_1(x-y) = \frac{1}{a} (-1)^{(x-y)/a},
\end{equation}
does not decay at all at large distances $x-y$ and thus describes an extremely
nonlocal action. For $d \geq 2$ the chirally symmetric perfect action remains 
nonlocal with a power-law decay of the couplings at large distances.

Although the nonlocality of the perfect action arose naturally by blocking the
theory from the continuum, from a practical point of view it is very
inconvenient. For example, in a numerical simulation it would be very demanding
to include couplings to far away neighbors. It follows from the 
Nielsen-Ninomiya theorem that, in order to obtain a local perfect action, one 
must break chiral symmetry explicitly. This can be done by modifying the above 
way of blocking from the continuum which was chirally covariant. If one chooses
to break chiral symmetry explicitly in the blocking procedure, the resulting 
perfect lattice action is not manifestly chirally invariant, but it is local.
This can be achieved by constructing a perfect lattice action $S[\Psibar,\Psi]$
as
\begin{eqnarray}
\exp(- S[\Psibar,\Psi])&=&\int {\cal D}\psibar {\cal D}\psi 
\exp\{- \frac{1}{(2 \pi)^d} \int d^dp \ 
\psibar(-p)[i \gamma_\mu p_\mu + m] \psi(p)\} \nonumber \\
&\times&\exp\{- \frac{1}{c} \frac{1}{(2 \pi)^d} \int_B d^dp \
[\Psibar(-p) - \sum_{n \in \Z^d} \psibar(- p - 2 \pi n/a) 
\Pi(p + 2 \pi n/a)] \nonumber \\ 
&\times&[\Psi(p) - \sum_{l \in \Z^d} \psi(p + 2 \pi l/a) 
\Pi(p + 2 \pi l/a)].
\end{eqnarray}
The coefficient $c$ is a source of explicit chiral symmetry breaking, which is
injected into the theory via the renormalization group transformation that maps
the continuum theory to the lattice theory. For $c \rightarrow 0$ we recover 
the chirally invariant but nonlocal perfect lattice action from before. In
general one obtains
\begin{equation}
\label{perfectprop}
\langle \Psibar(-p) \Psi(p) \rangle = 
\sum_{l \in \Z^d} [i \gamma_\mu (p_\mu + 2 \pi l_\mu/a) + m]^{-1} 
\Pi(p + 2 \pi l/a)^2 + c,
\end{equation}
which corresponds to a local perfect action as long as $c \neq 0$.

Let us vary $c$ in order to optimize the locality of the perfect action. For
this purpose we again consider $d = 1$. Then the sum in eq.(\ref{perfectprop}) 
can be performed analytically and the fermion propagator takes the form
\begin{equation}
\langle \Psibar(-p) \Psi(p) \rangle = \frac{1}{m} -
\frac{2}{m^2 a} [\coth(\frac{ma}{2}) - i \cot(\frac{pa}{2})]^{-1} + c.
\end{equation}
If we choose
\begin{equation}
c = \frac{\exp(m a) - 1 - m a}{m^2 a},
\end{equation}
the propagator reduces to
\begin{equation}
\langle \Psibar(-p) \Psi(p) \rangle = 
\left(\frac{\exp(m a) - 1}{m a}\right)^2 
\left[i \frac{1}{a} \sin(p a) + \frac{\exp(m a) - 1}{a} + 
\frac{2}{a} \sin^2(\frac{p a}{2})\right]^{-1}.
\end{equation}
This corresponds to the standard Wilson fermion action except that the mass 
$m$ is now replaced by $(\exp(m a) - 1)/a$. Hence, for the above choice of $c$,
in one dimension the perfect action is ultralocal, i.e.\ it has only 
nearest-neighbor interactions. In the massless limit $m = 0$ the optimal choice
for locality is $c = a/2$. When we go to more than one dimension the action 
remains local, but it is no longer ultralocal.

Next we derive the energy-momentum dispersion relation of perfect lattice
fermions. The fermion 2-point function takes the form
\begin{eqnarray}
\langle \Psibar(- \vec p,0) \Psi(\vec p,x_d) \rangle&=&\frac{1}{2 \pi} 
\int_{- \pi/a}^{\pi/a} dp_d \ \langle \Psibar(-p) \Psi(p) \rangle 
\exp(i p_d x_d) \nonumber \\
&=&\frac{1}{2 \pi} \int_{- \pi/a}^{\pi/a} dp_d \ \left\{
\sum_{l \in \Z^d} [i \gamma_\mu (p_\mu + 2 \pi l_\mu/a) + m]^{-1} 
\Pi(p + 2 \pi l/a)^2 + c \right\} \exp(i p_d x_d) \nonumber \\
&=&\frac{1}{2 \pi} \int_{- \infty}^{\infty} dp_d \
\sum_{l \in \Z^{d-1}} \frac{m}{(\vec p + 2 \pi \vec l/a)^2 + p_d^2 + m^2}
\nonumber \\
&\times&\prod_{i=1}^{d-1} 
\left(\frac{2 \sin(p_i a/2)}{p_i a + 2 \pi l_i}\right)^2
\left(\frac{2 \sin(p_d a/2)}{p_d a}\right)^2 \exp(i p_d x_d) + 
c \ \delta_{x_d,0} \nonumber \\
&=&\sum_{\vec l \in \Z^{d-1}} C(\vec p + 2 \pi \vec l/a) 
\exp(- E(\vec p + 2 \pi \vec l/a) x_d) + c \ \delta_{x_d,0}.
\end{eqnarray}
The sum over $l_d$ has been combined with the integral of $p_d$ over
$[- \pi/a,\pi/a]$ to an integral over the momentum space of the continuum
theory. The sum over the spatial $\vec l \in \Z^{d-1}$ leads to infinitely many
poles of the integrand, and hence to infinitely many states that contribute an
exponential to the 2-point function. The energies of these states are given by
the location of the poles, $E(\vec p + 2 \pi \vec l/a) = - i p_d$, with
\begin{equation}
E(\vec p + 2 \pi \vec l/a)^2 = - p_d^2 = 
(\vec p + 2 \pi \vec l/a)^2 + m^2.
\end{equation}
Hence, the energy-momentum dispersion relation of perfect lattice fermions is
exactly the same as in the continuum. In particular, as a result of exact
blocking from the continuum, there are no lattice artifacts. Furthermore, the 
form of the renormalization group blocking transformation has no effect on the 
physical spectrum. In particular, the explicit chiral symmetry breaking term 
proportional to $c$ only leads to a contact term $c \ \delta_{x_d,0}$ in the 
2-point function. Hence, it has no effect on the spectrum which is extracted 
from the 2-point function at large Euclidean time separations $x_d$.
Remarkably, the spectrum of the lattice theory displays the consequences of 
Poincar\'e invariance despite the fact that the lattice action only has the 
discrete lattice symmetries. 

Chiral symmetry is hidden in a similar way. Due to the explicit chiral symmetry
breaking parameter $c$ in the renormalization group blocking transformation,
even for $m = 0$ the perfect lattice action is not manifestly chirally 
invariant. Still, all physical consequences of chiral symmetry are correctly
reproduced by the perfect action. As we will see later, this is due to the
by now famous Ginsparg-Wilson relation \cite{Gin82},
\begin{equation}
\label{GWfree}
\{D^{-1},\gamma_5\} = a \gamma_5.
\end{equation}
Here $D$ is the lattice Dirac operator and $D^{-1}$ is the lattice fermion
propagator. Indeed, using the optimal parameter $c = a/2$ for the perfect 
action one obtains
\begin{equation}
\{D^{-1},\gamma_5\} = 
\{\sum_{l \in \Z^d} [i \gamma_\mu (p_\mu + 2 \pi l_\mu/a)]^{-1} 
\Pi(p + 2 \pi l/a)^2 + c,\gamma_5\} = 2 c \gamma_5 = a \gamma_5.
\end{equation}
The Ginsparg-Wilson relation is the key to understanding chiral symmetry on the
lattice \cite{Hasenfratz:1998jp}. In the continuum, chiral symmetry 
implies $\{D^{-1},\gamma_5\} = 0$.
If one insists on this relation also on the lattice, i.e.\ if one insists on
manifest chiral symmetry for a lattice action, the Nielsen-Ninomiya theorem 
implies fermion doubling (or, even worse, a violation of locality). The 
Ginsparg-Wilson relation $\{D^{-1},\gamma_5\} = a \gamma_5$ reduces to the 
relation $\{D^{-1},\gamma_5\} = 0$ in the continuum limit $a \rightarrow 0$. 
Still, at finite lattice spacing $a$, the right-hand side of the 
Ginsparg-Wilson relation implies an explicit breaking of chiral symmetry. In 
the case of the perfect action the explicit breaking is due to the parameter
$c = a/2$ in the renormalization group blocking transformation. This minimal 
explicit violation of chiral symmetry is sufficient to evade the 
Nielsen-Ninomiya theorem, and thus to avoid fermion doubling. Still, as we have
seen explicitly for the perfect action, the physics (for example, the spectrum)
remains the same as in the continuum. We will see later that the 
Ginsparg-Wilson relation leads to a natural definition of chiral symmetry on 
the lattice which reduces to the usual one in the continuum limit.

\section{Lattice QCD}

While many properties of lattice chiral symmetry can be studied in the free
theory, one certainly also needs to understand it in the interacting theory. 
Hence, it is now time to endow the quarks with their nontrivial QCD dynamics by
coupling them to the gluon field. However, before doing so, we first discuss
lattice Yang-Mills theory without quarks.

\subsection{\it Lattice Yang-Mills Theory}

Maintaining manifest gauge invariance is essential when gauge theories are
regularized on the lattice. In the continuum, gauge transformations involve
space-time derivatives of group-valued functions $\Omega(x)$. On the lattice
there are no infinitesimally close points, and continuum derivatives are 
usually simply replaced by finite differences. However, in order to maintain 
gauge invariance, one must proceed more carefully. Wegner and Wilson, as well 
as Smit, independently introduced the concept of a parallel transporter
$U_{x,\mu} \in SU(N_c)$ connecting neighboring lattice points $x$ and 
$x + \hat\mu$. The parallel transporter is related to an underlying continuum
gauge field $A_\mu(x) = i g A^a_\mu(x) T^a$ by
\begin{equation}
U_{x,\mu} = {\cal P} \exp \int_0^a dt \ A_\mu(x+\hat\mu t),
\end{equation}
where ${\cal P}$ denotes path-ordering. Under a non-Abelian gauge 
transformation the parallel transporter transforms as
\begin{equation}
U_{x,\mu}' = \Omega_x U_{x,\mu} \Omega_{x+\hat\mu}^\dagger.
\end{equation}
Wilson has constructed the Yang-Mills part of a simple lattice QCD action by 
multiplying parallel transporters around an elementary plaquette. The standard
Wilson action is constructed as a sum over all plaquettes
\begin{equation}
S_{YM}[U] = - a^4 \sum_{x,\mu,\nu} \frac{1}{g^2 a^2} \mbox{Tr}
[U_{x,\mu} U_{x+\hat\mu,\nu} U_{x+\hat\nu,\mu}^\dagger U_{x,\nu}^\dagger +
U_{x,\nu} U_{x+\hat\nu,\mu} U_{x+\hat\mu,\nu}^\dagger U_{x,\mu}^\dagger].
\end{equation}
It reduces to the continuum Yang-Mills action in the limit $a \rightarrow 0$.

To fully define the path integral we must also consider the measure. The 
lattice functional integral is obtained as an integral over all configurations 
of parallel transporters $U_{x,\mu}$, i.e.
\begin{equation}
Z = \prod_{x,\mu} \int_{SU(N_c)} dU_{x,\mu} \ \exp(- S_{YM}[U]).
\end{equation}
One integrates independently over all link variables using the local Haar 
measure $dU_{\mu,x}$ for each parallel transporter. The Haar measure is a left-
and right-invariant measure, i.e.
\begin{equation}
\int_{SU(N_c)} dU \ f(\Omega U) = \int_{SU(N_c)} dU \ f(U \Omega) = 
\int_{SU(N_c)} dU \ f(U),
\end{equation}
for any function $f(U)$ and for any $SU(N_c)$ matrix $\Omega$. It is convenient
to normalize the measure such that
\begin{equation}
\int_{SU(N_c)} dU = 1.
\end{equation}
For compact groups like $SU(N_c)$ the integration is over a finite domain. This
makes it unnecessary to fix the gauge in lattice QCD because the functional 
integral is finite even without gauge fixing. This is another important
advantage of the formulation using parallel transporters.

The Yang-Mills functional integral from above contains a single parameter ---
the bare gauge coupling $g$. When one wants to perform the continuum limit,
one must search for values of $g$ for which the correlation length of the
lattice theory diverges in lattice units. In the language of statistical 
mechanics one is looking for a second order phase transition. Due to asymptotic
freedom, in lattice QCD one expects a second order phase transition at 
$g \rightarrow 0$. To analyze the phase structure of a gauge theory one needs 
to study order parameters. A simple local order parameter like 
$\langle U_{x,\mu} \rangle$ is not useful. This follows from Elitzur's theorem
\cite{Eli75} which states that gauge-variant observables simply vanish. A 
useful order parameter in a gauge theory must be gauge invariant and, in 
addition, nonlocal. In a pure gluon theory a good order parameter was suggested
independently by Wegner and Wilson as
\begin{equation}
W_{\cal C} = \mbox{Tr} \prod_{(x,\mu) \in {\cal C}} U_{x,\mu}.
\end{equation}
For a rectangular curve ${\cal C}$ with side lengths $R$ and $T$ the Wilson 
loop describes the instantaneous creation and annihilation of a static 
quark-anti-quark pair at distance $R$ which then exists for a time $T$. The 
Wilson loop is related to the static quark-anti-quark potential $V(R)$ by
\begin{equation}
\lim_{T \rightarrow \infty} \langle W_{\cal C} \rangle \sim
\exp(- V(R) T).
\end{equation}
In QCD we expect quarks and anti-quarks to be confined to one another by
a potential rising linearly at large separations $R$, i.e.
\begin{equation}
\lim_{R \rightarrow \infty} V(R) \sim \sigma R,
\end{equation}
where $\sigma$ is the string tension. In a confinement phase the Wilson loop
hence shows an area law
\begin{equation}
\lim_{R,T \rightarrow \infty} \langle W_{\cal C} \rangle \sim
\exp(- \sigma R T).
\end{equation}
Confinement is indeed verified very accurately in numerical simulations of
lattice Yang-Mills theories.

\subsection{\it Standard Wilson Action for Lattice QCD}

We still need to couple the quarks to the gluons. First we do this by gauging 
$SU(N_c)$ in the action of free Wilson fermions of eq.(\ref{freewilson})
\begin{eqnarray}
\label{intwilson}
S_{QCD}[\overline \Psi,\Psi,U]&=&a^4 \sum_{x,\mu} \frac{1}{2a} 
(\Psibar_x \gamma_\mu U_{x,\mu} \Psi_{x+\hat\mu} -
\Psibar_{x+\hat\mu} \gamma_\mu U_{x,\mu}^\dagger \Psi_x) + 
a^4 \sum_x m \Psibar_x \Psi_x \nonumber \\
&+&a^4 \sum_{x,\mu} \frac{1}{2a} (2 \Psibar_x \Psi_x -
\Psibar_x U_{x,\mu} \Psi_{x+\hat\mu} -
\Psibar_{x+\hat\mu} U_{x,\mu}^\dagger \Psi_x)
\nonumber \\
&-&a^4 \sum_{x,\mu,\nu} \frac{1}{g^2 a^2} \mbox{Tr}
[U_{x,\mu} U_{x+\hat\mu,\nu} U_{x+\hat\nu,\mu}^\dagger U_{x,\nu}^\dagger +
U_{x,\nu} U_{x+\hat\nu,\mu} U_{x+\hat\mu,\nu}^\dagger U_{x,\mu}^\dagger].
\end{eqnarray}
In order to eliminate the doubler fermions we have introduced the Wilson term 
which breaks chiral symmetry explicitly. The lattice regularized functional 
integral takes the form
\begin{equation}
Z = \prod_x \int d\Psibar_x d\Psi_x \prod_{x,\mu} \int_{SU(N_c)} dU_{x,\mu} 
\exp(- S_{QCD}[\Psibar,\Psi,U]).
\end{equation}
It depends on two parameters --- the bare gauge coupling $g$ and the bare 
quark mass $m$. Due to asymptotic freedom, in order to reach the continuum 
limit one must take $g \rightarrow 0$. When one puts $m = 0$ for free Wilson 
fermions one reaches the chiral limit. In the interacting theory, on the other 
hand, this is no longer the case. In particular, since chiral symmetry is 
explicitly broken, the bare quark mass $m$ must be fine-tuned in order to reach
a massless limit. This fine-tuning is very unnatural from a theoretical point
of view. In particular, following the discussion in the introduction, if one 
imagines Wilson's lattice QCD as an oversimplified model for the short-distance
physics at the Planck scale, one could not understand at all why there are 
light fermions in Nature. The fine-tuning of $m$ is also inconvenient from a 
practical point of view. For example, in a numerical simulation of Wilson's 
lattice QCD one must fine-tune $m$ to many digits accuracy in order to make the
pion massless. If one does this at relatively large $g$, i.e.\ before one 
reaches the continuum limit $g \rightarrow 0$, this massless ``pion'' is not
even a proper Goldstone boson of a spontaneously broken chiral symmetry. For 
Wilson fermions an exact chiral symmetry that can break spontaneously does not 
exist at finite lattice spacing. It emerges only in the continuum limit after a
delicate fine-tuning of $m$.

\subsection{\it Ginsparg-Wilson Relation and L\"uscher's Lattice Chiral 
Symmetry}

\label{gwr}

In the discussion of the perfect free fermion action we have encountered the
Ginsparg-Wilson relation eq.(\ref{GWfree}). Following L\"uscher \cite{Lue98}, 
we will now use this relation to construct a version of chiral symmetry 
that is natural for
a lattice theory and reduces to the usual one in the continuum limit. For this
purpose we consider a lattice fermion action
\begin{equation}
S[\Psibar,\Psi,U] = a^4 \ \Psibar D[U] \Psi =
a^4 \sum_{x,y} \Psibar_x D[U]_{x,y} \Psi_y,
\end{equation}
which is defined in terms of the lattice Dirac operator $D[U]$. This 
operator should be local (i.e.\ it should decay exponentially at large 
distances $x-y$) but not ultralocal. The lattice Dirac operator obeys the
Ginsparg-Wilson relation if the corresponding fermion propagator 
$D[U]^{-1}$ satisfies
\begin{equation}
\{D[U]^{-1},\gamma_5\} = 
D[U]^{-1} \gamma_5 + \gamma_5 D[U]^{-1} = a \gamma_5.
\end{equation}
Alternatively, the Ginsparg-Wilson relation can be written as
\begin{equation}
\label{GWfull}
\gamma_5 D[U] + D[U] \gamma_5 = a D[U] \gamma_5 D[U].
\end{equation}
It is nontrivial to construct lattice actions that obey the Ginsparg-Wilson
relation. Until now we have seen that the perfect action for massless free
fermions indeed satisfies this relation. In the next two subsections we will 
see that the same is true for perfect actions for massless QCD as well as for 
overlap fermions. For the moment we don't worry about the concrete form of
$D[U]$, we just assume that it obeys eq.(\ref{GWfull}).

Let us first consider an infinitesimal chiral rotation of the form familiar 
from the continuum
\begin{eqnarray}
&&\Psi' = \Psi + \delta \Psi = (1 + i \varepsilon^a T^a \gamma_5) \Psi, 
\nonumber \\
&&\Psibar' = \Psibar + \delta \Psibar = 
\Psibar (1 + i \varepsilon^a T^a \gamma_5).
\end{eqnarray}
Here $T^a$ (with $a \in \{1,2,...,N_f^2 - 1\}$) are the generators of $SU(N_f)$
and $\varepsilon^a$ is a small parameter. In order to discuss flavor-singlet 
axial transformations with an infinitesimal parameter $\varepsilon^0$ we also 
define $T^0 = \1$. If the lattice action is local and has no fermion doubling, 
the Nielsen-Ninomiya theorem implies that it cannot be invariant under the 
above chiral rotations. On the other hand, the lattice fermion measure is 
invariant under the full chiral symmetry $U(N_f)_L \otimes U(N_f)_R$. This is 
very different from massless QCD in the continuum. In the continuum the action 
is invariant under $U(N_f)_L \otimes U(N_f)_R$ chiral transformations, while 
the measure is invariant only under $SU(N_f)_L \otimes SU(N_f)_R \otimes 
U(1)_{L=R}$. In particular, due to the anomaly the measure of the continuum 
theory is not invariant under flavor-singlet axial transformations, while the 
measure of the lattice theory is invariant.

Next we consider L\"uscher's modification of the standard chiral transformation
\begin{eqnarray}
\label{Lue}
&&\Psi' = \Psi + \delta \Psi = 
\left(1 + i \varepsilon^a T^a \gamma_5(1 - \frac{a}{2} D[U])\right) \Psi, 
\nonumber \\
&&\Psibar' = \Psibar + \delta \Psibar = 
\Psibar \left(1 + i \varepsilon^a T^a (1 - \frac{a}{2} D[U])\gamma_5\right).
\end{eqnarray}
Through $D[U]$ L\"uscher's lattice version of a chiral transformation 
depends on the gluon field. Still, in the continuum limit $a \rightarrow 0$ it
reduces to the standard chiral symmetry of the continuum theory. It is 
remarkable that eq.(\ref{Lue}) is a symmetry of any lattice action that obeys 
the Ginsparg-Wilson relation eq.(\ref{GWfull}). This follows from
\begin{eqnarray}
\Psibar' D[U] \Psi'&=&\Psibar \left(1 + i \varepsilon^a T^a
(1 - \frac{a}{2} D[U])\gamma_5\right) 
D[U] \left(1 + i \varepsilon^a T^a \gamma_5(1 - \frac{a}{2} D[U])\right) \Psi 
\nonumber \\
&=&\Psibar D[U] \Psi + \Psibar \left(i \varepsilon^a T^a 
[\gamma_5 D[U] + D[U] \gamma_5 - a D[U] \gamma_5 D[U]]\right) \Psi + 
O(\varepsilon^2) \nonumber \\
&=&\Psibar D[U] \Psi + O(\varepsilon^2).
\end{eqnarray}
Similarly, the variation of the lattice fermion measure takes the form
\begin{eqnarray}
\label{measure}
{\cal D}\Psibar' {\cal D}\Psi'&=&
{\cal D}\Psibar \left(1 - i \varepsilon^a T^a 
(1 - \frac{a}{2} D[U])\gamma_5\right) \left(1 - i \varepsilon^a T^a \gamma_5
(1 - \frac{a}{2} D[U])\right) {\cal D}\Psi \nonumber \\
&=&{\cal D}\Psibar {\cal D}\Psi \left(1 - i \varepsilon^a 
\mbox{Tr}[T^a \gamma_5 (2 - a D[U])]\right) + O(\varepsilon^2)
\nonumber \\
&=&{\cal D}\Psibar {\cal D}\Psi \left(1 + i \varepsilon^0 a
\mbox{Tr}[\gamma_5 D[U]]\right) + O(\varepsilon^2).
\end{eqnarray}
Hence, while any Ginsparg-Wilson fermion action is invariant under L\"uscher's 
lattice chiral symmetry, the lattice fermion measure is not. Exactly as in the 
continuum, the fermionic measure of the lattice theory changes under 
flavor-singlet axial transformations, while it is invariant under 
$SU(N_f)_L \otimes SU(N_f)_R \otimes U(1)_{L=R}$. We will see later that the
non-invariance of the fermionic measure under flavor-singlet axial 
transformations indeed gives rise to the correct axial anomaly.

\subsection{\it Classically Perfect Action for Lattice QCD}

Hasenfratz and Niedermayer have initiated and carried out an impressive program
of explicitly constructing nonperturbative lattice actions that are perfect at
least at the classical level \cite{HN94}. The pure gauge part of a classically 
perfect action is the fixed point of a renormalization group blocking 
transformation that maps lattice gauge fields $U$ on a fine lattice with 
lattice spacing $a$ to lattice gauge fields $U'$ on a coarser lattice with 
spacing $2 a$. Denoting the blocking kernel by $T_{YM}[U,U']$, the classically 
perfect action obeys the condition
\begin{equation}
\label{classperf}
S_{YM}[U'] = \min_U \left(S_{YM}[U] + T_{YM}[U,U']\right),
\end{equation}
which implicitly defines $S[U]$. This equation can be solved numerically by an
iterative procedure on a multi-layer of finer and finer lattices. 

The fermionic part of a classically perfect massless fermion action results as 
a fixed point of a renormalization group transformation which maps a fermion
field $\Psi$ on the fine lattice to a fermion field $\Psi'$ on the coarse 
lattice using a blocking kernel $T_F[U]$, i.e.
\begin{equation}
\Psi' = T_F[U] \Psi \ \Rightarrow \ \Psi'_{x'} = \sum_x T_F[U]_{x',x} \Psi_x.
\end{equation}
Here $x$ is a point on the fine lattice and $x'$ is a point on the coarse
lattice. The fermionic part of the classically perfect action is quadratic in 
the fermion fields and can be written as
\begin{equation}
S_F[\Psibar,\Psi,U] = a^4 \ \Psibar D_P[U] \Psi = 
a^4 \ \sum_{x,y} \Psibar_x D_P[U]_{x,y} \Psi_y.
\end{equation}
The corresponding fermion propagator $D_P[U]^{-1}$ obeys the equation
\begin{equation}
D_P[U']^{-1} = T_F[U] D_P[U]^{-1} T_F[U] + c \ \Rightarrow
D_P[U']^{-1}_{x',y'} = \sum_{x,y} T_F[U]_{x',x} D_P[U]^{-1}_{x,y} T_F[U]_{y,y'}
+ c \ \delta_{x',y'}.
\end{equation}
Here $U$ is the gauge field on the fine lattice that minimizes the expression
on the right-hand side of eq.(\ref{classperf}) given a gauge field $U'$ on the 
coarse lattice. The parameter $c$ is analogous to the one introduced in the 
perfect action of the free theory. For an appropriate fermionic blocking kernel
$T_F[U]$ and a choice of $c$ that again optimizes locality one can show that 
the classically perfect action obeys the Ginsparg-Wilson relation. 

There are a number of important properties of the classically perfect
fixed point action. For example,
\begin{itemize}
\item[(i)] The free gauge and fermion parts of the action reproduce
the exact relativistic spectrum of the continuum.
\item[(ii)] It is possible to define a topological charge such that the 
action of any gauge field configuration with topological charge $Q$ is larger 
than $8\pi^2 |Q|/g^2$, a relation that is known from the continuum theory.
\item[(iii)] There is an exact index theorem on the lattice that relates 
the fermionic zero modes of the fixed point Dirac operator and the 
topological charge of the lattice gauge field.
\item[(iv)] Due to the Ginsparg-Wilson relation there is an exact chiral
symmetry. 
\end{itemize}
For more details we refer the reader to the article by Hasenfratz 
\cite{Has98} and references therein.

Although classically perfect actions satisfy a number of attractive 
properties, they are difficult to implement in practice. A highly nontrivial 
and crucial step is to find practical (but still accurate) parameterizations 
of the perfect action. Although the initial work involved spin and gauge 
models in two dimensions, recently practical approximations of the fixed point 
actions for lattice QCD have emerged. In the pure gauge theory a useful 
parametrization was found in \cite{DeGrand:1995ji}, while the fermionic 
problem was tackled in \cite{Hasenfratz:2000xz}. The resulting approximations 
of classically perfect actions yield physical results that are almost 
completely free of lattice artifacts \cite{DeG95,Has02}.

\subsection{\it Domain Wall and Overlap Fermions}

\label{dwov}

In the early nineties Kaplan proposed a novel method to preserve chirality 
on the lattice \cite{Kap92}. The idea was to use the fact that chiral 
fermions become trapped on domain walls \cite{Cal85}. Kaplan used a 
Wilson-Dirac operator in five dimensions with a mass term that is a function 
of the fifth direction. In particular, the mass term changes sign creating a 
four-dimensional domain wall at the points where it vanishes. A 
four-dimensional chiral fermion is then trapped on the domain wall. In the 
meantime Narayanan and Neuberger were developing an idea of using an infinite 
number of regulator ``flavor'' fields to preserve chirality 
\cite{Narayanan:wx}. 
They realized that Kaplan's construction was equivalent to
their idea since the fifth dimension is analogous to a flavor space. They used 
their interpretation and argued that the determinant of a chiral fermion in the
background of a gauge field is equivalent to the overlap of two many-body 
fermionic ground states \cite{Narayanan:sk}. Initially, it seemed that the 
overlap was a reliable technique to regulate even chiral gauge theories on a 
lattice. Unfortunately, it was realized that the fermions with opposite 
chirality (which originate from an anti-wall) cannot be easily 
decoupled \cite{Golterman:1993th,Golterman:1994at}. 

Although it was not clear whether the domain wall and overlap approach
gave a completely satisfactory formulation of lattice chiral gauge theories
involving no doublers of opposite chirality, something highly nontrivial had 
been achieved. It was possible to construct a lattice theory with a fermion 
with opposite chirality and use it effectively to preserve chiral symmetry 
in a vector-like gauge theory. An elegant way to use a five-dimensional 
fermion action to represent quarks in lattice QCD was first proposed by 
Shamir \cite{Shamir:1993zy} and elaborated further by Furman and Shamir in 
\cite{Furman:ky}. This fermion is commonly referred to as a domain wall 
fermion and is used 
extensively in lattice simulations. Its action is constructed on a 
five-dimensional space-time lattice with coordinates $(x,x_5)$, where $x$ 
refers to the usual four dimensions and $x_5 \in \{a_5,2 a_5,...,L_5\}$ refers 
to the fifth direction of finite extent $L_5$. Since the fifth direction is 
physically different from the other directions we have introduced a new lattice
spacing $a_5$ in that direction. The domain wall fermion action is given by
\begin{equation}
S_F[\Psibar,\Psi,U] = a^4 a_5 \sum_{x,x_5,y,y_5} 
\Psibar_{x,x_5} D_{DW}[U]_{x,x_5;y,y_5} \Psi_{y,y_5}.
\end{equation}
The domain wall Dirac operator is given by
\begin{eqnarray}
D_{DW}[U]_{x,x_5;y,y_5}&=&\delta_{x_5,y_5} D^\parallel[U]_{x,y} +
\delta_{x,y} D^\perp[U]_{x_5,y_5}, \nonumber \\
D^\parallel[U]_{x,y}&=&M \delta_{x,y} + 
\sum_\mu \frac{1}{2a} \left(\gamma_\mu U_{x,\mu} \delta_{x+\hat{\mu},y} - 
\gamma_\mu U^\dagger_{x-\hat\mu,\mu} \delta_{x-\hat\mu,y}\right) \nonumber \\
&-&\sum_\mu \frac{1}{2 a} \left(2 \delta_{x,y} 
- U_{x,\mu} \delta_{x+\hat{\mu},y} 
- U^\dagger_{x-\hat\mu,\mu} \delta_{x-\hat\mu,y}\right), \nonumber \\
D^\perp[U]_{x_5,y_5}&=&\left\{\begin{array}{ll}
(P_R\, \delta_{2 a_5,y_5} - \delta_{a_5,y_5})/a_5 - m P_L\, 
\delta_{L_5,y_5} & \mbox{for} \ x_5 = 1, \\
(P_R\, \delta_{x_5+a_5,y_5} + P_L\, \delta_{x_5-a_5,y_5} - 
\delta_{x_5,y_5})/a_5 & \mbox{for} \ a_5 < s < L_5, \\
(P_L\, \delta_{L_5-a_5,y_5} - \delta_{L_5,y_5})/a_5 - m P_R\, 
\delta_{a_5,y_5} & \mbox{for} \ x_5 = L_5. \end{array}\right.
\end{eqnarray}
Here $P_R$ and $P_L$ are the chiral projection operators defined in 
eq.(\ref{projectors}). In the above action the parameter $M$ is not the 
mass of the quark that is bound to the wall. By comparing with Wilson fermions 
one sees that the sign of the Wilson term has changed. In order to produce 
massless quarks one should set $0 \leq M a_5 \leq 2$ at tree level 
\cite{Kap92} and take $L_5 \rightarrow \infty$. There is a technical 
problem that needs to be taken into account. When $L_5$ becomes infinite there 
are only $N_f$ flavors of four-dimensional massless quarks bound to the wall,
but there is an infinite number of modes at the cut-off. This may cause 
spurious effects at low energies. Hence, one needs to use bosonic 
(Pauli-Villars
type) fields to cancel the contribution of these high-energy modes. For a 
detailed discussion of how this can be accomplished we refer the reader to 
\cite{Furman:ky}.

There is a close connection between the domain wall approach and the overlap 
formula developed by Narayanan and Neuberger. Neuberger realized that it is 
possible to find an analytic formula for an effective Dirac operator that 
describes the massless chiral mode of the domain wall fermion. Using his 
insight on the overlap formula for vector-like gauge theories 
\cite{Narayanan:1994gw}, he found a simple and elegant formula for the 
four-dimensional Dirac operator \cite{Neuberger:1997fp}, which is referred to 
as the overlap Dirac operator and which is given by
\begin{equation}
D_O[U] = \frac{1}{2 a} \Big[1 + \gamma_5 \frac{H[U]}{\sqrt{H[U]^2}}\Big]
\label{ovD}
\end{equation}
where $H[U] = \gamma_5 D^\parallel[U]$ and $D^\parallel[U]$ is the 
operator we defined above in the context of the domain wall fermion. In 
order to obtain massless quarks one needs to set $0\leq M a_5 \leq 2$ as 
before. In fact, it is possible to find an analytic formula for an effective 
Dirac operator that represents the chiral massless modes of the domain wall 
fermion even for finite $L_5$ 
\cite{Narayanan:1994gw,Neu98a,Kikukawa:1999sy,Giusti:2002rx}.
This operator takes the form
\begin{equation}
D_{L_5}[U] = \frac{1}{a} \Big[1 + \gamma_5 
\frac{(1 + \tilde{H}[U])^{L_5/a_5} - (1 - \tilde{H}[U])^{L_5/a_5}}
{(1 + \tilde{H}[U])^{L_5/a_5} + (1 - \tilde{H}[U])^{L_5/a_5}}\Big],
\end{equation}
where
\begin{equation}
\tilde{H}[U] = \gamma_5 \tilde{X}[U], \ 
\tilde{X}[U] = \frac{a_5 D^\parallel[U]}{2 + a_5 D^\parallel[U]}.
\end{equation}
In the limit of $L_5 \rightarrow \infty$ one obtains
\begin{equation}
\lim_{L_5 \rightarrow \infty} D_{L_5}[U] = D_{DWO}[U] = \frac{1}{2 a} 
\Big[1 + \gamma_5 \frac{\tilde{H}[U]}{\sqrt{\tilde{H}[U]^2}}\Big],
\end{equation}
which reduces to the overlap Dirac operator $D_O[U]$ when 
$a_5 \rightarrow 0$. 

Although today we know that the Ginsparg-Wilson relation leads to an exact 
chiral symmetry on the lattice, this connection was not appreciated until 
recently. After Ginsparg and Wilson discovered this interesting relation they 
found it difficult to explicitly construct a local operator that satisfies
it. The relation was soon forgotten. After the discovery of 
perfect and overlap fermions the relation was rediscovered by Hasenfratz. It is
straightforward to check that $D_O[U]$ and $D_{DWO}[U]$ indeed satisfy the 
Ginsparg-Wilson relation. These new Dirac operators couple every pair of sites 
on the lattice. It is possible to show that if one wants to benefit from good 
chiral properties of Ginsparg-Wilson fermions, one has to give up the notion of 
ultralocal actions \cite{Horvath:1998cm}. However, as has been shown in 
\cite{Hernandez:1998et}, close to the continuum limit the couplings in the 
overlap Dirac operator fall off exponentially with the distance. In this
sense these new Dirac operators are still local. Unfortunately, the closeness
to the continuum limit is quite important to maintain both the chiral and
locality properties of the Dirac operator. Recently, a physical picture based 
on the locality of zero modes of the Dirac operator was used to map out the 
regions in coupling constant space where $D_O[U]$ and $D_{DWO}[U]$ lead to a 
good regularization of massless quarks \cite{Golterman:2003qe}.

There have been recent efforts to generalize the Ginsparg-Wilson relation 
\cite{Fujikawa:2000my} and use this as a guide to construct new classes of 
Dirac operators \cite{Fujikawa:2001av}. Since Dirac operators which satisfy 
the Ginsparg-Wilson relation exactly are computationally very demanding, 
there have also been efforts to find perfect Dirac operators that satisfy 
the Ginsparg-Wilson fermions approximately \cite{Bietenholz:1998ut}. Another
approach has been to expand the most general lattice Dirac operator
in a basis of simple operators. The coefficients of the expansion then are
determined using the Ginsparg-Wilson relation \cite{Gattringer:2000js}.
This approach has been used to construct a practical operator for lattice
simulations \cite{Gattringer:2000qu}.

\subsection{\it Quenched Approximation}

The chiral limit of lattice QCD is computationally very demanding because
all algorithms contain a step in which the inverse of the Dirac operator in 
a fixed gauge field background needs to be computed. If the Dirac operator 
has small eigenvalues this step becomes very time consuming. Further, 
in the presence of small eigenvalues the number of inversions necessary 
before a statistically independent configuration is generated also 
increases leading to inefficiencies in the algorithms.

The full partition function of lattice QCD is given by
\begin{eqnarray}
Z &=& \prod_x \int d\Psibar_x d\Psi_x 
\prod_{x,\mu} \int_{SU(N_c)} dU_{x,\mu} \ \exp(- S_{YM}[U])\ 
\exp(-a^d \Psibar D[U] \Psi) \nonumber \\
&=& \prod_{x,\mu} \int_{SU(N_c)} dU_{x,\mu} \ \exp(- S_{YM}[U])\ 
\mathrm{det} D[U].
\end{eqnarray}
Observables take the form
\begin{equation}
\langle O \rangle = \frac{1}{Z} \prod_{x,\mu} \int_{SU(N_c)} dU_{x,\mu} \ 
\exp(- S_{YM}[U]) \ \mathrm{det} D[U] \ O[U],
\end{equation}
where $O[U]$ is the observable in the background gauge field $U$. For
example, when the observable is a fermion bilinear, $O[U]$ is constructed
from a quark propagator. The difficulties in the chiral limit arise due 
to the determinant factors in the above relations. In order to avoid such 
difficulties, the determinant factor is sometimes dropped. This is an 
approximation in which one ignores the effects of virtual quark loops, 
which is commonly referred to as the quenched approximation. It should be 
pointed out that the quenched approximation is not a systematically controlled 
one. It is possible to take into account some effects of virtual quark loops
by allowing the quarks in the loop to be heavier. This is the so-called
partially quenched approximation. Many calculations today use either a
quenched or partially quenched approximation to compute physical quantities 
for light quarks.

Not surprisingly, in the chiral limit the quenched approximation introduces 
unphysical singularities. These singularities are again related to the small 
eigenvalues of the Dirac operator. However, now they enter observables through
the quark propagator. Using quenched chiral perturbation theory, one can 
predict the form of these singularities in various observables close to the 
chiral limit \cite{Bernard:1992mk,Bernard:1993sv}.
For example, the chiral condensate is expected to diverge logarithmically
with a dimensionless coefficient $\delta$. Such spurious divergences are 
referred to as quenched chiral logarithms. Recently, several groups have 
evaluated $\delta$ with results varying from about $0.03$ 
to about $0.26$. For further details we refer the reader to 
\cite{Wittig:2002ux}. 

The quenched approximation is particularly severe in the case of Wilson 
fermions close to the chiral limit. This is because, due to the absence of
an exact chiral symmetry, the Wilson Dirac operator may contain arbitrarily 
small eigenvalues when the quark mass is small even if it is non-zero. 
In the case of staggered fermions and Ginsparg-Wilson fermions one can show 
that the eigenvalues of the Dirac operator cannot be 
smaller than the quark mass, as expected in the continuum. This problem 
hinders simulations with dynamical light Wilson quarks.

By adding a so-called twisted mass term to the two-flavor Wilson fermion 
action, it is possible to regulate the small eigenvalues of the Dirac 
operator \cite{Bardeen:1998dd}. This term does not change the continuum 
limit of the theory, since one can argue that --- provided the continuum 
limit is taken carefully --- the extra term is equivalent to a redefinition 
of the fermion field \cite{Frezzotti:2000nk}. This new approach to lattice
QCD with an additional twisted mass term is becoming another practical 
approach to study the effects of light quarks in QCD, which has been tested in 
\cite{Jansen:2003ir}.

\section{Special Features of Ginsparg-Wilson Fermions}

We have already discussed in section \ref{gwr} how the Ginsparg-Wilson relation
leads to a new realization of chiral symmetry in a finite lattice theory. This 
lattice chiral symmetry makes domain wall and overlap fermions, which were
introduced in section \ref{dwov}, special in various ways. In this section we 
discuss some of the features of these Ginsparg-Wilson fermions that allow us to
relate lattice quantities to continuum physical quantities more directly than 
it was possible before.

\subsection{\it Anomaly and Topological Charge on the Lattice}

It is well-known that the flavor-singlet chiral symmetry is anomalous in QCD. 
This can be shown in the continuum in essentially two ways: (i) By taking 
suitable care of the ultraviolet regulator in the diagrammatic approach one can
show that the divergence of the flavor-singlet axial current is equal to 
the topological charge density of the gluon field; (ii) in the functional
integral representation of QCD the fermionic integration measure is not 
invariant under flavor-singlet axial transformations when the Dirac operator in
the background of a gluon field configuration has a non-zero index 
\cite{Fujikawa:1979ay,Fujikawa:1980eg}. Before the discovery of the 
Ginsparg-Wilson relation and its consequences, in a lattice formulation the 
anomaly could only be understood after a complex calculation 
\cite{Karsten:1978nb}. Since the fermionic measure was always invariant under 
chiral transformations, either the anomalous symmetry was explicitly broken by 
the lattice action (as in the case of Wilson or staggered fermions) or the 
fermion doubling would cancel the anomaly completely (as for naive fermions). 
The only viable approach was to calculate the appropriate diagrams in the 
lattice theory, which is usually quite tedious, and then to take the 
continuum limit.

The discovery of the exact chiral symmetry of Ginsparg-Wilson fermions allows 
one to derive the anomaly in a straightforward fashion as explained 
in \cite{Lue98}. Here we review the essential steps of that proof. 
Consider the expectation value of an operator ${\cal O}[\Psibar,\Psi]$ in
the background of a gauge field $U$ given by
\begin{equation}
\langle {\cal O}\rangle = \prod_x \int d\Psibar_x d\Psi_x \ 
{\cal O}[\Psibar,\Psi] \exp(- a^4 \sum_{x,y} \Psibar_x D_{x,y}[U] \Psi_y),
\end{equation}
where $D[U]_{x,y}$ is a lattice Dirac operator that obeys the 
Ginsparg-Wilson relation. For simplicity we have suppressed color, flavor, and 
Dirac indices. Using the variation of the fermionic measure eq.(\ref{measure})
under L\"uscher's infinitesimal flavor-singlet axial transformations as well
as the invariance of the action one can show that
\begin{equation}
\langle \delta{\cal O} \rangle =  \varepsilon^0 a\ \mbox{Tr}[\gamma_5 D[U]]
\langle {\cal O} \rangle.
\end{equation}
As argued in \cite{Lue98}, if $D[U]$ obeys the Ginsparg-Wilson 
relation it is possible to show that
\begin{equation}
a (z - D[U]) \gamma_5 (z - D[U]) = z(2 - az) \gamma_5
- (1 - a z) [(z - D[U]) \gamma_5 + \gamma_5(z - D[U])],
\end{equation}
where $z$ is a complex number not contained in the spectrum of $D[U]$.
After multiplying both sides of this equation from the right with 
$(z - D[U])^{-1}$ and taking the trace one gets
\begin{equation}
- a \mbox{Tr}[\gamma_5 D[U]] = z(2 - az) \mbox{Tr}[\gamma_5 (z - D[U])^{-1}].
\end{equation}
We can now divide both sides by the factor $z(2 - az)$ and integrate over a 
small circle centered at the origin that does not encircle any spectral value 
of $D[U]$ other than $0$. Since
\begin{equation}
P_0 = \oint \frac{dz}{2 \pi i} \ (z - D[U])^{-1}
\end{equation}
projects on the subspace of zero modes of $D[U]$, we then get
\begin{equation}
a \mbox{Tr}[\gamma_5 D[U]] = 2 \ (n_- - n_+) = 2 \ \mbox{index}(D[U]).
\label{gwindex}
\end{equation}
where $n_\pm$ represent the number of zero modes of $D[U]$ which are
also eigenstates of $\gamma_5$ with eigenvalues $\pm 1$. This yields 
the anomalous Ward-identity
\begin{equation}
\label{awid}
\langle \delta{\cal O} \rangle =  2 \ \mbox{index}(D[U]) 
\langle {\cal O} \rangle,
\end{equation}
which is familiar from the continuum but which is now valid in a completely 
regularized finite lattice theory.

Often a lattice Dirac operator $D[U]$ is $\gamma_5$-Hermitean, i.e.\ it obeys 
the relation $D^\dagger = \gamma_5 D \gamma_5$, in addition to satisfying the 
Ginsparg-Wilson
relation. In that case one can simplify the above discussion which relates 
$a\mbox{Tr}[\gamma_5 D[U]]$ to the index of $D[U]$. This was done in 
\cite{Lal98,Niedermayer:1998bi} where the lattice index theorem was first
derived. One can show that the
eigenstates $u_\lambda$ of $a D[U]$ are such that the eigenvalues $\lambda$
fall on a circle given by $\lambda = 1 - \mathrm{e}^{i\alpha}$. There are 
three types of eigenstates:
(i) Those with $\lambda=0$ such that $\gamma_5 u_\lambda = \pm u_\lambda$. Let
$n_\pm$ represent the number of such eigenstates with eigenvalue $\pm 1$;
(ii) those with $\lambda=2$ such that $\gamma_5 u_\lambda = \pm u_\lambda$. Let
$n'_\pm$ represent the number of such eigenstates with eigenvalue $\pm 1$;
(iii) those with complex (non-real) $\lambda$, in which case 
$\gamma_5 u_\lambda = u_{\lambda^*}$. Since $\mbox{Tr}(\gamma_5) = 0$,
we must have $n_- - n_+ = n'_+-n'_-$. Using these relations it is easy to 
derive eq.(\ref{gwindex}).

The above calculation shows that there is indeed an anomaly in a theory
with Ginsparg-Wilson fermions. However, it is also important to show that 
the anomaly reproduces the well-known results in the continuum limit.
This has been shown perturbatively for a class of overlap Dirac operators
in \cite{Rei99} and nonperturbatively for the overlap Dirac operator 
of eq.(\ref{ovD}) in \cite{Kik99,Fuj99,Suz99,Adam02}. More recently, the 
nonperturbative arguments were extended to the overlap Dirac operator 
that is constructed using a perfect fermion action \cite{Adams:2003hy}. 
Thus, from a variety of studies we can conclude with confidence that 
the above calculation does indeed reproduce the correct anomaly in the 
continuum limit.

In the continuum, the index of the Dirac operator is related to the topological
charge of the background gauge field configuration. On the lattice there is no 
unique way to define the topological charge. A naive discretization of the 
topological charge does not even yield an integer on the lattice. A geometric
construction of the topological charge is possible which then leads to integer 
values \cite{Luscher:1981zq}. The index of a Ginsparg-Wilson Dirac operator is 
yet another definition of the topological charge. In this case we define the 
topological charge as
\begin{equation}
\label{topch}
Q = \frac{1}{2} a \mbox{Tr}[\gamma_5 D[U]].
\end{equation}
This definition is more meaningful since it is the same charge that enters the 
anomalous Ward identity in eq.(\ref{awid}). As we will see in the 
next section, the Witten-Veneziano mass formula can be derived on the lattice 
using this definition of the topological charge. Further discussions of the 
anomaly and the topological charge can be found in a recent review 
\cite{Niedermayer:1998bi} and the references therein.

\subsection{\it Witten-Veneziano Mass Formula on the Lattice}

The mass of the $\eta'$-meson is closely connected to the anomalous 
flavor-singlet chiral symmetry of QCD. In the absence of the anomaly one can 
argue that the flavor-singlet chiral symmetry would be spontaneously broken 
leading to a massless Goldstone boson for massless quarks. The absence of a 
low-mass flavor-singlet pseudo-scalar particle in Nature with the correct 
flavor quantum numbers is usually attributed to the anomaly. To compute this 
mass from first principles is one of the outstanding challenges in QCD. Due to 
the nonperturbative nature of the physics involved, the only viable approach to
this problem from first principles is using lattice QCD.

Interestingly, the physics of the $\eta'$-meson mass can be understood more 
easily
in certain limits of QCD. For example, in the limit $N_c \rightarrow \infty$,
with $g^2 N_c$ and $N_f$ held fixed \cite{Witten:1979vv}, or by assuming that 
the anomalous flavor-singlet axial Ward identities retain their 
validity order by order in an expansion in $u \equiv N_f/N_c$ around $u = 0$ 
\cite{Veneziano:1979ec}. In both cases one can derive a leading-order 
Witten-Veneziano relation for the mass of the $\eta'$-meson which is given by
\begin{equation}
\label{etam}
m_{\eta'}^2 = \frac{2N_f}{F_\pi^2} \int d^4x \ \langle q(x) q(0) \rangle_{YM},
\end{equation}
where $F_\pi$ is the pion decay constant and $q(x)$ is the topological 
charge density
\begin{equation}
\label{Q}
q(x) = - \frac{1}{32\pi^2} \epsilon_{\mu\nu\rho\sigma}
\mbox{Tr}[F_{\mu\nu}(x) F_{\rho\sigma}(x)].
\end{equation}
Here the trace is over color indices. The subscript $YM$ in eq.(\ref{etam}) 
indicates that the $qq$-correlation function is to be computed in the pure 
Yang-Mills theory, i.e.\ in the absence of quarks.

Although the Witten-Veneziano relation is valid only in certain limits of QCD, 
one can use it to estimate $m_{\eta'}$ in a lattice QCD calculation. 
Unfortunately, eq.(\ref{etam}) is rather formal and cannot be translated to 
lattice QCD without addressing a number of subtleties \cite{Alles:1996nm}
\footnote{For a recent review we refer the reader to \cite{Giusti:2001xh} and 
the references therein. Here we review only some basic points.}. Two main 
problems need to be solved in order to make eq.(\ref{etam}) rigorous and of 
practical use. One has to: (i) find a properly normalized lattice definition
of the topological charge density $q(x)$; (ii) subtract from $q(x)q(0)$ 
appropriate
contact terms, so as to define it properly in the continuum limit. As we have 
discussed earlier, different lattice definitions of $q(x)$ are possible. 
Furthermore, the second problem can be quite subtle since the subtraction may
involve contact terms of the form $c \ \delta(x)$ which contribute a finite 
constant to the $\eta'$-meson mass. Some of these subtleties were first
pointed out in \cite{Sei87}. The reader is also referred to a recent
discussion in \cite{Sei02}.

One of the special features of Ginsparg-Wilson fermions is that we can use 
eq.(\ref{topch}) to define the topological charge density $q_x$ (which obeys
$a^4 \sum_x q_x = Q$) as
\begin{equation}
q_x = \frac{1}{2 a^3} \mbox{Tr}[\gamma_5 D[U]_{x,x}].
\end{equation}
In this expression the trace is only over the color and Dirac indices, but not
over the space-time index $x$. As explained in \cite{Giusti:2001xh}, this 
definition allows one to derive
\begin{equation}
\label{etamlat}
m_{\eta'}^2 = \frac{2N_f}{F_\pi^2} a^4 \sum_x \langle q_0 q_x \rangle_{YM},
\end{equation}
in the appropriate limit in lattice QCD. Thus one obtains a relation for the 
$\eta'$-meson mass exactly as in the continuum. Thanks to the chiral symmetry 
properties of Ginsparg-Wilson fermions, no additional subtractions are 
necessary.

\subsection{\it Renormalization of Operators}

Computing the matrix elements of fermionic currents in QCD is necessary for
determining a variety of physical quantities such as hadronic decay constants, 
electromagnetic and weak form factors, quark masses, etc. A knowledge of the
lattice renormalization constants of these operators is necessary to relate 
the matrix elements computed using lattice simulations to the corresponding 
ones defined in continuum renormalization schemes, like the $\overline{MS}$
scheme, used in experimental data analysis. Recently, several groups have 
contributed to this subject. We refer the reader to 
\cite{Sint00} and references therein for a recent review of renormalization in
lattice field theory. The existence of exact symmetries simplifies the 
computations of these renormalization constants enormously. Hasenfratz has
discussed operator renormalization in the context of perfect actions 
\cite{Has98a}. In this section, using a few examples, we illustrate how one
can obtain constraints among various renormalization constants. A more general 
method to renormalize massless Ginsparg-Wilson fermions in lattice gauge 
theories has been worked out in \cite{Reisz:1999ck}. For an extensive 
discussion of perturbative renormalization, especially in the context of 
Ginsparg-Wilson fermions we refer the reader to the recent review 
\cite{Capitani:2002mp} and references therein.

Consider the renormalization constants for local bilinear quark operators of 
the form
\begin{equation}
\label{operators}
{\cal O}_\alpha(x) = \psibar(x) \Gamma_\alpha \psi(x),
\end{equation}
where $\Gamma_\alpha$ denotes generic Dirac matrices, i.e.\ 
$1,\,\gamma_5,\,\gamma_\mu,\,\gamma_\mu\gamma_5,\,\sigma_{\mu\nu}$. Specific 
bilinear operators, denoted according to their Lorentz group transformations, 
are 
\begin{equation}
S(x) = \psibar(x) \psi(x),\ P(x) = \psibar(x) \gamma_5 \psi(x),\
V^a_\mu(x) = \psibar(x) \gamma_\mu T^a \psi(x),\ 
A^a_\mu(x) = \psibar(x) \gamma_\mu \gamma_5 T^a \psi(x).
\end{equation}
The lattice renormalization constants can be obtained by the equation
\begin{equation}
\Gamma_{\cal O}^{\overline{MS}}(p,\mu) = Z_{\cal O}(a \mu) Z_\psi(a \mu) 
\Gamma_{\cal O}^L(p,a),
\label{zo}
\end{equation}
where $\Gamma_{\cal O}^L(p,a)$ and $\Gamma_{\cal O}^{\overline{MS}}(p,\mu)$ are
the two-quark one-particle irreducible vertex functions with an insertion of 
${\cal O}$ calculated respectively on the lattice and in the continuum using 
the $\overline{MS}$ renormalization scheme, and $Z_\psi(a\mu)$ is the wave 
function renormalization constant. In the continuum, chiral symmetry imposes 
constraints among the various renormalization constants. For example, it is 
well-known that $Z_S = Z_P$ and $Z_V = Z_A$. Such relations reduce the amount 
of work in computing the renormalization constants. On the lattice one often 
has to deal with mixing of operators with lower-dimensional operators leading 
to additional subtractions that may diverge with an inverse power of the 
lattice spacing. Fortunately, with Ginsparg-Wilson fermions the renormalization
of operators is also quite analogous to the continuum. For example, one can
argue that the relations $Z_S = Z_P$ and $Z_V = Z_A$ remain valid 
\cite{Alexandrou:2000kj}.

In order to see the usefulness of the Ginsparg-Wilson relation, let us define 
a new set of fermion bilinear operators
\begin{equation}
\label{opprimes}
{\cal O}'_{\alpha} = 
\Psibar \Gamma_\alpha \left(1 -\frac{a}{2} D[U]\right) \Psi,
\end{equation}
where $D[U]$ is the Ginsparg-Wilson Dirac operator appearing in the 
fermionic action of the theory. The addition of the term depending on 
$D[U]$ is harmless since its contributions vanish in the continuum limit. 
This leads to the fact that the renormalization constants $Z_{\cal O'}$ for the
new operators are the same as $Z_{\cal O}$. In fact, these new ``primed'' 
operators are 
more natural in a Ginsparg-Wilson theory at finite lattice spacings. As pointed
out in \cite{Chandrasekharan:1998wg} and \cite{Kikukawa:1998py}, for finite 
lattice spacing the spontaneous breaking of non-singlet chiral symmetry is 
related to the vacuum expectation value of the operator $S'$, i.e.\ it occurs 
if $\langle S' \rangle \neq 0$. Further, by using the infinitesimal 
flavor-singlet chiral transformations given in eq.(\ref{Lue}) one can show
that
\begin{equation}
\label{trspp}
\delta S'(x) = 2 \varepsilon^0 P'(x), \ \delta P'(x) = 2 \varepsilon^0 S'(x).
\end{equation}
Under the non-singlet chiral transformations the vector and axial currents 
$V'^a_\mu$ and $A'^a_\mu$ transform as
\begin{equation}
\delta V'^a_\mu = i f^{abc} \varepsilon^b A'^c_\mu, \
\delta A'^a_\mu = i f^{abc} \varepsilon^b V'^c_\mu.
\end{equation}
Using the corresponding Ward identities one can show that  
\begin{equation}
\label{zpr}
Z_{S'} = Z_{P'}, Z_{V'} = Z_{A'}.
\end{equation}
Thus, we find that $Z_S = Z_P$ and $Z_V = Z_A$. 

There is also a simplification in determining the renormalization of the quark 
mass using Ginsparg-Wilson fermions when the bare fermion mass is introduced in
the action by writing the massive Dirac operator as 
\begin{equation}
D_m[U] = D[U] + m \left(1 - \frac{a}{2} D[U]\right).
\end{equation}
The renormalization of the fermion mass is then related to that of the operator
$S'$ denoted by $Z_S$. In particular, it is easy to show that
\begin{equation}
Z_m = Z^{-1}_S.
\end{equation}
The renormalized quark mass is then obtained as $m_{ren} = Z_m m$. 

Using these ideas, one-loop perturbative computations of these renormalization 
constants were performed with overlap fermions in \cite{Alexandrou:2000kj}, 
while a nonperturbative approach has been used in \cite{Hernandez:2001yn} to 
determine the quark condensate and quark masses. Nonperturbative 
renormalization has also been studied extensively with domain wall fermions, 
the details of which can be found in \cite{Blum:2001sr}.

\subsection{\it Lattice Simulations with Ginsparg-Wilson Fermions}

Over the past few years, numerical simulations with overlap fermions, domain 
wall fermions, and approximately perfect fermions have begun to appear. 
Computationally, these fermions are about fifty to a hundred times more 
expensive than conventional fermions. Thus, the most reliable results from 
lattice simulations using these new fermions are still obtained in the quenched
approximation where the fermion determinant is ignored. In this section we 
review some of the recent work.

Simulations with domain wall fermions are in much better shape thanks to both a
larger number of researchers working on this subject and faster computers being
used in the computations. A detailed study of the chiral properties of domain 
wall quarks can be found in \cite{Blum:2000kn} and \cite{AliKhan:2000iv}. 
Recently, an extensive calculation of quenched QCD with light quarks using 
overlap fermions was performed in \cite{Chiu:2003iw}. Chiral properties of 
overlap fermions were studied earlier in \cite{DeGrand:2002gm,Dong:2001fm}. The
first results using approximately classically perfect fermions, which obey the 
Ginsparg-Wilson relation to a good accuracy, have also been published in 
\cite{Gattringer:2003qx}.

Calculations using domain wall fermions typically use lattices of size 
$24^3 \times 40$ to $32^3 \times 60$ at a lattice cut-off which varies from 
$1/a = 2$ GeV to 3 GeV. On the other hand, simulations with overlap fermions
are currently done on lattices of sizes from $12^3 \times 24$, $16^3 \times 32$
to $20^4$ at lattice cut-offs of $1/a = 1.33$ GeV to 2 GeV. Results using 
approximately perfect fermions have been obtained on $8^3\times 24$, 
$12^3 \times 24$, and $16^3\times 32$ lattices at lattice cut-offs 
$1/a = 1.3$ GeV, 2 GeV, and 2.5 GeV.

A variety of quantities can be calculated in lattice simulations. This includes
the hadron spectrum, the topological susceptibility, the quark condensate, the 
pion decay constant, the average of the up and down quark masses 
$\overline{m} = (m_u + m_d)/2$, the strange quark mass $m_s$, weak matrix 
elements like the kaon $B_K$ parameter, properties of the nucleon, such as its 
axial charge $g_A$ and the parameters that arise in quenched and unquenched
chiral perturbation theory. A comparison of the results obtained in different 
studies indicates that there are systematic errors in many of these quantities 
that are presently of the order of $10-20\%$. For example, the topological 
susceptibility defined as
\begin{equation}
\chi_t = \frac{\langle Q^2 \rangle}{V}
\end{equation}
is found to vary from about $(0.176 \mbox{GeV})^4$ in \cite{Chiu:2003iw},
$(0.196 \mbox{GeV})^4$ in \cite{Hasenfratz:2001qp}, to 
$(0.213 \mbox{GeV})^4$ in \cite{DeGrand:2002gm}. Finite temperature effects
were recently studied in \cite{Gattringer:2002mr} and it was found that
$\chi_t$ varies from about $(0.191\mbox{GeV})^4$ at $T=0.88T_c$ to
$(0.100\mbox{GeV})^4$ at $T=1.31T_c$ where $T_c$ is the deconfinement
temperature. Currently, several groups are performing a 
careful analysis of the systematic errors due to finite volume effects,
finite lattice spacing errors, as well as the effects of dynamical quark loops.

Another important parameter in QCD is the chiral condensate. However, 
unlike $\chi_t$ this quantity needs a proper definition in a 
renormalization scheme \cite{Hernandez:1999cu}. For Ginsparg-Wilson 
fermions a nonperturbative renormalization group invariant definition 
of the condensate was introduced in \cite{Hernandez:2001yn}. Using this 
definition the condensate was found to be $(0.243(10) \mbox{GeV})^3$ 
in \cite{Hernandez:2001hq} using overlap fermions and 
$(0.235(11) \mbox{GeV})^3$ in \cite{Hasenfratz:2002rp} using 
the perfect action approach. When these are converted to the 
$\overline{\mathrm{MS}}$ scheme at a scale of $2 \mbox{GeV}$ one obtains
$(0.266(15) \mbox{GeV})^3$, which is consistent with 
$(0.250(3) \mbox{GeV})^3$ obtained in \cite{Chiu:2003iw}. These values 
are in agreement with the value $(0.256(8) \mbox{GeV})^3$ obtained 
in \cite{Blum:2000kn} using domain wall fermions. The errors in the
chiral condensate seem to be at the $4-8\%$ level.

One of the most impressive achievements in lattice QCD is the determination of 
the hadron spectrum. It is indeed exciting that one can compute the masses
of a variety of hadrons from first principles and compare them with
experiments. In the quenched approximation this spectrum shows 
about $10\%$ deviations from the experimental results \cite{Aoki:1999}. These
deviations appear to reduce when one introduces dynamical quarks 
\cite{AliKhan:2000mw}. Since most calculations are not close to the continuum 
limit, to some extent the deviations depend on the quantity used to determine 
the lattice spacing. Recently, results using 
chirally improved fermions in the quenched limit have begun to emerge. 
Results using fixed point fermions suggest that deviations from the 
experimental results are strongly correlated with their experimentally 
observed widths \cite{Gattringer:2003qx}. In other words, unstable 
particles are more severely affected by quenching than stable ones. Using
domain wall fermions, the quenched spectrum has been investigated 
in \cite{Aoki:2002vt}.

Several groups have recently obtained results on kaon physics 
using overlap and domain wall fermions. The main goal in these projects is to 
extract the kaon $B_K$ parameter, which plays a central role in understanding 
$CP$ violation in the kaon system. The value of $B_K$ depends on the 
renormalization scale and scheme. When evaluated at $\mu = 2$ GeV in the 
$\overline{MS}$ scheme the value appears to be around $0.57$ with an error of
about $5-10\%$. We refer the reader to the work in 
\cite{DeGrand:2003in,Garron:2003cb,AliKhan:2001wr,Blum:2001xb} for further 
details. Quark masses have also been studied in 
\cite{Chiu:2003iw,Giusti:2001pk} using overlap fermions and in 
\cite{AliKhan:2001wr,Blum:1999xi} using domain wall fermions. The current 
estimates are $\overline{m} = 3.5 - 4$ MeV and $m_s = 99 - 133$ MeV. 

Recently, domain wall quarks have been used to compute the nucleon axial charge
$g_A$. One such calculation finds $g_A = 1.21(5)$ in the chiral limit. 
This should be compared with the experimental result $g_A = 1.267(3)$. We refer
the reader to \cite{Sasaki:2003jh} for further details and for a comparison 
with earlier work. There are also interesting results on the excited states of 
the nucleon from different types of chiral fermion actions
\cite{Sasaki:2001nf,Brommel:2003jm,Dong:2003zf}. Finally, we like to
mention that efforts to compute the parton distribution functions from lattice
QCD are well-developed (see \cite{Negele:2002vs} for a recent review).
Recently, a method to measure even generalized parton distribution functions 
has been proposed \cite{Hagler:2003jd} and has been applied using domain wall 
fermions \cite{Schroers:2003mf}. Such calculations have also been performed
using Wilson fermions \cite{Gok96,Gok01,Gok04}.

There are many other studies we have not touched on in this article. All these 
studies indicate that, with sufficient computing power, at least in the 
quenched approximation, computations using Ginsparg-Wilson fermions are 
feasible. Dynamical fermion calculations are also being envisioned. However,
in the 
absence of an algorithmic breakthrough for dynamical fermions, it is likely 
that most computations will be limited to quark masses that are still too large
for chiral perturbation theory to work reliably. This may keep us from studying
the full dynamics of chiral symmetry from first principles for some time.

\section{Low-Energy Effective Theories}

Due to spontaneous chiral symmetry breaking, at low energies the QCD dynamics
is dominated by Goldstone bosons \cite{Gol61} --- the pions for $N_f = 2$. 
It is possible to use a low-energy effective description that only involves 
the Goldstone boson fields \cite{Col69,Cal69,Wei79}. Chiral perturbation 
theory provides a systematic low-energy expansion 
that predicts the pion dynamics based on symmetry principles and a few 
low-energy parameters (like the pion decay constant $F_\pi$ and the chiral 
order parameter $\langle \psibar \psi \rangle$) whose values can be determined 
either from experiments or from lattice QCD calculations \cite{Gas84}. The 
low-energy interactions of pions and nucleons can be understood in baryon 
chiral perturbation theory \cite{Geo84a,Gas88,Jen91,Ber92,Bec99}. Effective 
theories that are relevant to more than one nucleon can be useful even in
the absence of pions \cite{Kap98,Bed98}. For a recent review of this 
subject we refer the reader to \cite{Bed02}. 

Usually, low-energy effective theories are treated perturbatively directly in 
the continuum. In order to investigate nonperturbative effects that may arise 
within the effective theory, it is interesting to also regularize it on 
the lattice \cite{Myi93,Smi99,Lew01}. There are efforts underway to understand
nuclear matter on the lattice \cite{Mue00,Che03,Lee03} starting from some 
effective theory. Interestingly, the nonlinear realization of chiral symmetry 
on the lattice does not lead to the same subtleties (like the fermion doubling 
problem) that one faces within the microscopic QCD theory 
\cite{Chandrasekharan:2003wy}.

\subsection{\it Effective Theory for Goldstone Bosons}

The Goldstone bosons are the lightest particles in QCD. Therefore they dominate
the dynamics of the strong interactions at low energies. It is possible to use 
a low-energy effective description that only involves the Goldstone boson 
fields. This is not only true for QCD but also for any other system with a 
continuous global symmetry $G$ which is spontaneously broken to a subgroup $H$.
The Goldstone bosons are described by fields in the coset space $G/H$ in which
points are identified if they are connected by symmetry transformations of the 
unbroken subgroup $H$. In QCD we have $G/H = SU(N_f)$. Hence the Goldstone 
boson fields are represented by special unitary matrices $U(x) \in SU(N_f)$. 
Under global chiral rotations they transform as
\begin{equation}
U(x)' = L U(x) R^\dagger.
\end{equation}
Goldstone bosons interact weakly at low energies. Their effective Lagrangian is
constructed as a derivative expansion. The leading term of the pion effective 
action takes the form
\begin{equation}
S[U] = \int d^4 x \ 
\left[\frac{F_\pi^2}{4} \mbox{Tr}(\p_\mu U^\dagger \p_\mu U)
+ \frac{\langle \psibar \psi \rangle}{2 N_f} 
\mbox{Tr}({\cal M} U^\dagger + U {\cal M}^\dagger)\right].
\end{equation}
The first term on the right-hand side is chirally invariant. Its prefactor is
the pion decay constant $F_\pi$ which determines the strength of the 
interaction between the Goldstone bosons. The second term is the chiral 
symmetry breaking mass term which contains the quark mass matrix. Under chiral 
transformations this term transforms as
\begin{equation}
\mbox{Tr}({\cal M} {U'}^\dagger + U' {\cal M}^\dagger) =
\mbox{Tr}({\cal M} R U^\dagger L^\dagger + L U R^\dagger {\cal M}^\dagger).
\end{equation}
If all quark masses are equal, i.e.\ if ${\cal M} = m \1$, the Lagrangian is 
invariant against $SU(N_f)_F$ flavor rotations for which $R=L$. For a general 
diagonal mass matrix the flavor symmetry is reduced to 
$\prod_{f=1}^{N_f} U(1)_f$. 

The constants $F_\pi$ and $\langle \psibar \psi \rangle$ determine the 
low-energy dynamics at leading order and enter the effective theory as a 
priori unknown parameters. These parameters can be determined from experiments 
or through nonperturbative lattice QCD calculations. Up to these two low-energy 
constants the Goldstone boson dynamics is completely determined by chiral 
symmetry. At higher energies additional terms arise in the effective theory. 
Again, they are restricted by chiral symmetry and they contain new low-energy
parameters --- the Gasser-Leutwyler coefficients. Chiral perturbation theory 
(with mass-degenerate quarks) is a systematic low-energy expansion around the 
classical vacuum configuration $U(x) = \1$. One writes
\begin{equation}
U(x) = \exp(2 i \pi^a(x) T^a/F_\pi), \ a \in \{1,2,...,N_f^2 - 1\},
\end{equation}
where $T^a$ are the generators of $SU(N_f)$, and one then expands in powers of
$\pi^a(x)$. In this way one can derive, for example, the Gellmann-Oakes-Renner 
relation
\begin{equation}
M_\pi^2 = \frac{2 m \langle \psibar \psi \rangle}{N_f F_\pi^2}.
\end{equation}

\subsection{\it Effective Theory for Nucleons and Pions}

Chiral perturbation theory can be extended to sectors with non-zero baryon 
number. Nucleons enter the low-energy effective theory in the form of a Dirac 
spinor field $\psi(x)$ and $\psibar(x)$ that transforms as an $SU(2)_I$
isospin doublet. Global chiral rotations $L \otimes R \in SU(2)_L \otimes 
SU(2)_R$ can be realized nonlinearly on this field using the transformations
\begin{equation}
\psi(x)' = V(x) \psi(x), \ \psibar(x)' = \psibar(x) V(x)^\dagger.
\end{equation}
The field $V(x)$ depends on $L$ and $R$ as well as on the field $U(x)$ and can
be written as
\begin{equation}
V(x) = R(R^\dagger L U(x))^{1/2} (U(x)^{1/2})^\dagger =
L(L^\dagger R U(x)^\dagger)^{1/2} U(x)^{1/2}.
\end{equation}
For transformations in the unbroken isospin vector subgroup $SU(2)_I = 
SU(2)_{L=R}$ the field $V(x)$ reduces to the global flavor transformation 
$V(x) = L = R$. The local transformation $V(x)$ is a nonlinear representation 
of chiral symmetry which has the form of a local $SU(2)$ transformation, 
despite the fact that it represents just a global $SU(2)_L \otimes SU(2)_R$ 
symmetry. 

In order to construct a chirally invariant (i.e.\ $SU(2)$ ``gauge'' invariant) 
action one needs an $SU(2)$ flavor ``gauge'' field. For this purpose one 
constructs a field $u(x) \in SU(2)$ from the pion field $U(x)$ as
\begin{equation}
u(x) = U(x)^{1/2}.
\end{equation}
The matrix $u(x)$ is located in the middle of the shortest geodesic connecting 
$U(x)$ with the unit-matrix $\1$ in the group manifold of $SU(2)$. Under chiral
rotations the field $u(x)$ transforms as
\begin{equation}
u(x)' = L u(x) V(x)^\dagger = V(x) u(x) R^\dagger.
\end{equation}
The anti-Hermitean composite field
\begin{equation}
\label{v}
v_\mu(x) = \frac{1}{2}[u(x)^\dagger \partial_\mu u(x) + 
u(x) \partial_\mu u(x)^\dagger],
\end{equation}
transforms as a ``gauge'' field
\begin{equation}
v_\mu(x)' = \frac{1}{2}
[V(x) u(x)^\dagger L^\dagger \partial_\mu (L u(x) V(x)^\dagger) +
V(x) u(x) R^\dagger \partial_\mu (R u(x)^\dagger V(x)^\dagger)] = 
V(x) (v_\mu(x) + \partial_\mu) V(x)^\dagger.
\end{equation}
A Hermitean composite field is given by
\begin{equation}
\label{a}
a_\mu(x) = \frac{i}{2}[u(x)^\dagger \partial_\mu u(x) - 
u(x) \partial_\mu u(x)^\dagger],
\end{equation}
which transforms as
\begin{equation}
a_\mu(x)' = \frac{1}{2}
[V(x) u(x)^\dagger L^\dagger \partial_\mu (L u(x) V(x)^\dagger) -
V(x) u(x) R^\dagger \partial_\mu (R u(x)^\dagger V(x)^\dagger)] =
V(x) a_\mu(x) V(x)^\dagger.
\end{equation}

The leading terms in the Euclidean action of a low-energy effective theory for 
nucleons and pions take the form
\begin{eqnarray}
\label{action2}
S[\psibar,\psi,U]&=&\int d^4x \ \{M \psibar \psi + 
\psibar \gamma_\mu (\partial_\mu + v_\mu) \psi +
i g_A \psibar \gamma_\mu \gamma_5 a_\mu \psi \nonumber \\
&+&\frac{F_\pi^2}{4} \mbox{Tr}[\partial_\mu U^\dagger \partial_\mu U] -
\frac{\langle \overline \psi \psi \rangle}{4} \mbox{Tr}[{\cal M} U^\dagger +
{\cal M}^\dagger U]\}.
\end{eqnarray}
Here $M$ is the nucleon mass generated by spontaneous chiral symmetry breaking
and $g_A$ is the coupling to the isovector axial current. It is remarkable that
--- thanks to the nonlinear realization of chiral symmetry --- the fermion mass
term is chirally invariant. This makes sense because the mass $M$ arises from 
the spontaneous breakdown of chiral symmetry even in the chiral limit. It is
remarkable that fermions with a nonlinearly realized chiral symmetry do not 
contribute to anomalies \cite{Geo84a,Man84}. In the low-energy effective theory
anomalies enter through the Wess-Zumino-Witten term \cite{Wes71,Wit83}.

\subsection{\it Effective Theory for Constituent Quarks}

The chiral quark model of Georgi and Manohar \cite{Geo84b} is formulated in 
terms of gluons, pions, and constituent quarks $\psi(x)$ and 
$\psibar(x)$ which transform in the fundamental representations of $SU(N_c)$ 
and $SU(N_f)$. In particular, under the nonlinearly realized 
$SU(N_f)_L \otimes SU(N_f)_R$ chiral symmetry the constituent quark field 
transforms as
\begin{equation}
\psi(x)' = V(x) \psi(x), \ \psibar(x)' = \psibar V(x)^\dagger,
\end{equation}
while under an $SU(N_c)$ color gauge transformation
\begin{equation}
\psi(x)' = \Omega(x) \psi(x), \ \psibar(x)' = \psibar \Omega(x)^\dagger.
\end{equation}
The Euclidean action of the chiral quark model is given by
\begin{eqnarray}
S[\psibar,\psi,U,A]&=&\int d^4x \ \{M \psibar \psi + 
\psibar \gamma_\mu (\partial_\mu + v_\mu + A_\mu) \psi +
i g_A \psibar \gamma_\mu \gamma_5 a_\mu \psi \nonumber \\
&+&\frac{F_\pi^2}{4} \mbox{Tr}[\partial_\mu U^\dagger \partial_\mu U] -
\frac{1}{4} \langle \overline \psi \psi \rangle \mbox{Tr}[{\cal M} U^\dagger +
{\cal M}^\dagger U] - \frac{1}{2 g^2} \mbox{Tr}[F_{\mu\nu} F_{\mu\nu}]\}.
\end{eqnarray}
In this case, $M$ is the constituent quark mass which is generated by
spontaneous chiral symmetry breaking. It should be noted that in the effective 
theory the strong gauge coupling $g$ is weaker than in QCD itself. However, due
to the coupling to the gluon field the constituent quarks are still confined. 

The chiral quark model is based on the assumption that the energy scale for 
chiral symmetry breaking is larger than the one for confinement. An effective 
description in terms of constituent quarks which receive their mass from chiral
symmetry breaking before they get confined by residual low-energy gluons should
then make sense. In fact, the phenomenological success of the nonrelativistic 
quark model may suggest that this picture is indeed correct. In their work, 
Georgi and Manohar provided a framework that puts the nonrelativistic quark 
model on a solid field theoretical basis.

A potential problem of the chiral quark model is related to the confinement 
scale. The value of $g$ may be so small that there are unacceptably low-lying 
glueball states. For the same reason, there might be a low-temperature 
deconfinement phase transition in the gluon sector significantly below the 
finite temperature chiral phase transition. These potential problems are 
impossible to address quantitatively in the continuum formulation of the chiral
quark model because they involve the nonperturbative dynamics of confinement. 
In order to be able to address these issues, it is useful to formulate the 
constituent quark model on the lattice.

\subsection{\it Lattice Formulation of Nonlinearly Realized Chiral Symmetry}

Let us construct theories with a nonlinearly realized chiral symmetry on the
lattice. The Goldstone boson field $U_x \in SU(N_f)$ naturally lives on the 
sites $x$ of a four-dimensional hypercubic lattice and it transforms as
\begin{equation}
U_x' = L U_x R^\dagger,
\end{equation}
under global chiral rotations. As in the continuum, the field $u_x \in SU(N_f)$
is constructed as $u_x = U_x^{1/2}$ which transforms as
\begin{equation}
u_x' = L u_x V_x^\dagger = V_x u_x R^\dagger.
\end{equation}

We now proceed to the construction of the lattice analog $V_{x,\mu} \in 
SU(N_f)$ of the continuum flavor ``gauge'' field $v_\mu(x)$ which is a flavor 
parallel transporter along a lattice link in the group $SU(N_f)$. In analogy to
the continuum expression eq.(\ref{v}) for $v_\mu(x)$ we construct
\begin{equation}
\tilde V_{x,\mu} = \frac{1}{2a}[u_x^\dagger u_{x+\hat\mu} + 
u_x u_{x+\hat\mu}^\dagger].
\end{equation}
In the continuum limit $\tilde V_{x,\mu} = \exp[a v_\mu(x+\hat\mu/2)]$. 
However, at finite lattice spacing $\tilde V_{x,\mu}$ is in general not an 
element of $SU(N_f)$, just a complex $N_f \times N_f$ matrix in the group 
$GL(N_f)$. One can project a group-valued parallel transporter $V_{x,\mu} 
\in SU(N_f)$ out of $\tilde V_{x,\mu}$ by performing a $GL(N_f)/SU(N_f)$ coset 
decomposition \cite{Chandrasekharan:2003wy}. By construction, under the 
nonlinearly realized chiral symmetry $V_{x,\mu}$ then transforms as a 
parallel transporter, i.e.
\begin{equation}
\label{Vtrans}
V_{x,\mu}' = V_x V_{x,\mu} V_{x+\hat\mu}^\dagger.
\end{equation}

Next we construct a lattice version of the continuum field $a_\mu(x)$ defined 
in (\ref{a}). For this purpose, we first construct
\begin{equation}
\tilde A_{x,\mu} = \frac{i}{2a}[u_x^\dagger u_{x+\hat\mu} - 
u_x u_{x+\hat\mu}^\dagger].
\end{equation}
While in the continuum $a_\mu(x)' = V(x) a_\mu(x) V(x)^\dagger$, the lattice 
field $\tilde A_{x,\mu}$ transforms as
\begin{equation}
\tilde A_{x,\mu}' = V_x \tilde A_{x,\mu} V_{x+\hat\mu}^\dagger.
\end{equation}
Also, in contrast to the continuum field $a_\mu(x)$, the lattice field 
$\tilde A_{x,\mu}$ is in general neither traceless nor Hermitean. It is 
therefore more natural to introduce the field
\begin{equation}
\label{Aleft}
A^L_{x,\mu} = \frac{1}{2}[\tilde A_{x,\mu} V_{x,\mu}^\dagger +
V_{x,\mu} \tilde A_{x,\mu}^\dagger] - \frac{1}{2N_f} \mbox{Tr}
[\tilde A_{x,\mu} V_{x,\mu}^\dagger + V_{x,\mu} \tilde A_{x,\mu}^\dagger] \1,
\end{equation}
which, by construction, is traceless and Hermitean and which transforms as 
\begin{equation}
{A^L_{x,\mu}}' = V_x A^L_{x,\mu} V_x^\dagger,
\end{equation}
with the matrix $V_x$ located at the site $x$ on the left end of the link
$(x,\mu)$. Similarly, we define the object
\begin{equation}
\label{Aright}
A^R_{x,\mu} = \frac{1}{2}[V_{x,\mu}^\dagger \tilde A_{x,\mu} +
\tilde A_{x,\mu}^\dagger V_{x,\mu}] - \frac{1}{2N_f} \mbox{Tr}
[V_{x,\mu}^\dagger \tilde A_{x,\mu} + \tilde A_{x,\mu}^\dagger V_{x,\mu}] \1,
\end{equation}
which transforms as 
\begin{equation}
{A^R_{x,\mu}}' = V_{x+\hat\mu} A^R_{x,\mu} V_{x+\hat\mu}^\dagger,
\end{equation}
with the matrix $V_{x+\hat\mu}$ located at the site $x+\hat\mu$ on the right 
end of the link $(x,\mu)$. It should be noted that $A^R_{x,\mu}$ and
$A^L_{x,\mu}$ are not independent but are related by parallel transport, i.e.
\begin{equation}
A^R_{x,\mu} = V_{x,\mu}^\dagger A^L_{x,\mu} V_{x,\mu}.
\end{equation}

\subsection{\it Constituent Quarks on the Lattice}

Using the lattice construction of a nonlinearly realized chiral symmetry 
presented in the previous subsection, it is now straightforward to put, for
example, Georgi and Manohar's chiral quark model on the lattice. The resulting 
fermion action takes the form
\begin{eqnarray}
\label{GMaction}
S_F[\overline \Psi,\Psi,V,A,U]&=&a^4 \sum_x M \Psibar_x \Psi_x + 
a^4 \sum_{x,\mu} \frac{1}{2a} 
(\Psibar_x \gamma_\mu V_{x,\mu} U_{x,\mu} \Psi_{x+\hat\mu} -
\Psibar_{x+\hat\mu} \gamma_\mu V_{x,\mu}^\dagger U_{x,\mu}^\dagger \Psi_x) 
\nonumber \\
&+&a^4 \sum_{x,\mu} \frac{i g_A}{2a}
(\Psibar_x \gamma_\mu \gamma_5 A^L_{x,\mu} \Psi_{x} +
\Psibar_{x+\hat\mu} \gamma_\mu \gamma_5 A^R_{x,\mu} \Psi_{x+\hat\mu})
\nonumber \\
&+&a^4 \sum_{x,\mu} \frac{1}{2a} (2 \Psibar_x \Psi_x -
\Psibar_x V_{x,\mu} U_{x,\mu} \Psi_{x+\hat\mu} -
\Psibar_{x+\hat\mu} V_{x,\mu}^\dagger U_{x,\mu}^\dagger \Psi_x).
\end{eqnarray}
Here $U_{x,\mu} \in SU(N_c)$ denotes the standard Wilson color parallel 
transporters living on the lattice links. It is interesting to ask if the gluon
dynamics of QCD can be modeled successfully in the chiral quark model. Since 
the chiral quark model does not represent a systematic low-energy expansion of 
QCD, one should not expect to obtain quantitative results directly relevant to 
QCD. However, even if only qualitative insight into the success of the 
nonrelativistic quark model can be gained, this would be quite interesting.

The Wilson term in the above action removes the doubler fermions. As we 
discussed before, in standard lattice QCD this term breaks chiral symmetry 
explicitly. Remarkably, when chiral symmetry is nonlinearly realized, not only 
the fermion mass term (proportional to $M$) but also the Wilson term is 
chirally invariant. The only source of explicit chiral symmetry breaking is the
current quark mass matrix ${\cal M}$. It is interesting to ask how the 
Nielsen-Ninomiya theorem \cite{Nie81} has been avoided. Clearly, the action of
eq.(\ref{GMaction}) is local. The corresponding Dirac operator is given by
\begin{eqnarray}
\label{Dirac}
D[V,A,U]_{x,y}&=&M \delta_{x,y} + 
\sum_\mu \frac{1}{2a} (\gamma_\mu V_{x,\mu} U_{x,\mu} \delta_{x+\hat\mu,y} -
\gamma_\mu V_{x-\hat\mu,\mu}^\dagger U_{x-\hat\mu,\mu}^\dagger 
\delta_{x-\hat\mu,y}) \nonumber \\
&+&\sum_\mu \frac{i g_A}{2a} (\gamma_\mu \gamma_5 A^L_{x,\mu} \delta_{x,y} +
\gamma_\mu \gamma_5 A^R_{x-\hat\mu,\mu} \delta_{x,y}) \nonumber \\
&+&\sum_\mu \frac{1}{2a} 
(2 \delta_{x,y} - V_{x,\mu} U_{x,\mu} \delta_{x+\hat\mu,y} -
V_{x-\hat\mu,\mu}^\dagger U_{x-\hat\mu,\mu}^\dagger \delta_{x-\hat\mu,y}).
\end{eqnarray}
The Nielsen-Ninomiya theorem assumes that the Dirac operator anti-commutes with
$\gamma_5$. This is not the case when chiral symmetry is nonlinearly realized, 
and hence one of the basic assumptions of the Nielsen-Ninomiya theorem is not 
satisfied. Interestingly, Ginsparg-Wilson fermions evade the Nielsen-Ninomiya 
theorem by violating the same assumption. Of course, the nonlinear realization
of chiral symmetry requires an explicit pion field which is not present in the 
fundamental QCD Lagrangian. 

Remarkably, not only the lattice fermion action but also the lattice fermion 
measure is gauge invariant. At first sight this seems to be a severe problem 
because gauge invariance of both the fermion action and the fermion measure 
implies that lattice fermions with nonlinearly realized chiral symmetry do not 
contribute to anomalies. One might even suspect that the doubler fermions have 
not been properly removed and thus have canceled the anomalies of the physical 
fermions. Fortunately, this is not the case. Indeed fermions with a nonlinearly
realized chiral symmetry do not contribute to anomalies. Instead, as mentioned
in the continuum discussion, the anomalies are contained in the 
Wess-Zumino-Witten term \cite{Wes71,Wit83} which must be added to the
lattice action explicitly.

\section{Conclusions}

In this review we have discussed some basic issues of chiral symmetry on the
lattice. The recent lattice developments have put chiral symmetry on a solid 
theoretical basis at a nonperturbative level. As discussed in the introduction,
lattice QCD can now explain nonperturbatively why nucleons can exist 
naturally, i.e.\ without fine-tuning, far below the Planck scale, provided that
space-time has additional hidden dimensions. This in turn explains why gravity 
is so weak, a nontrivial result one obtains from lattice QCD without doing
any numerical work. Even chiral gauge theories like the standard model have now
been constructed rigorously beyond perturbation theory. This is a very 
substantial step forward in the theoretical formulation of the basic laws of 
Nature. The development of Ginsparg-Wilson lattice fermions is also beginning 
to revolutionize practical lattice QCD simulations. In particular, if new
algorithmic developments go hand in hand with the recent theoretical insights,
Ginsparg-Wilson fermions may lead to substantial progress towards an accurate
numerical solution of QCD. As usual, many new questions arise based on the new
insights. For example, supersymmetry still waits to be put on rigorously solid 
grounds beyond perturbation theory. Also many practical numerical calculations 
with Ginsparg-Wilson fermions still need to be done. We hope that we have 
provided a certain basis for newcomers to enter this very active field of 
current research.

\section*{Acknowledgements}

Over the years we have discussed chiral symmetry both in the continuum and on
the lattice with many colleagues including O.~B\"ar, T.~Bhattacharya, 
W.~Bietenholz, R.~Brower, G.~Colangelo, N.~Christ, E.~Farhi, J.~Gasser, 
M.~G\"ockeler, J.~Goldstone, M.~Golterman, P.~Hasenfratz, R.~Jackiw, R.~Jaffe, 
J.~Jersak, D.~Kaplan, A.~Kronfeld, M.~Laursen, H.~Leutwyler, M.~L\"uscher, 
A.~Manohar, T.~Mehen, J.~Negele, R.~Narayanan, H.~Neuberger, F.~Niedermayer, 
M.~Pepe, G.~Schierholz, Y.~Shamir, S.~Sharpe, J.~Smit, R.~Springer, F.~Steffen,
and F.~Wilczek. We gratefully acknowledge the insights they generously shared 
with us. This work was supported in part by funds provided by the U.S.\ 
Department of Energy (D.O.E.) under cooperative research agreement 
DE-FG02-96ER40945 and by the Schweizerischer Nationalfond (SNF).

\end{document}